 \documentclass[a4paper,11pt]{article}
  
\usepackage{jcappub} 
\usepackage[T1]{fontenc}
\usepackage{here}
\usepackage{siunitx}
\usepackage{ulem}
\usepackage[utf8]{inputenc}
\usepackage{subfigure}

\usepackage{dcolumn}
\usepackage{bm}
\usepackage{amsmath}
\usepackage{amsthm}
\usepackage{ascmac}
\usepackage{xparse} 
\usepackage{mathtools} 
\usepackage{blkarray} 
\usepackage{amscd,verbatim}
\usepackage{amssymb}
\usepackage[all]{xy}
\usepackage{tikz}
\usepackage{tikz-cd}
\usepackage{pgf,pgfplots}
\usepackage{enumitem}
\usepackage{slashed}

\pgfplotsset{compat=1.18} 

\usetikzlibrary{decorations.pathmorphing}

\newcommand\dd{\text{d}}

\begin{document}

\title{
Gravitational waves from metastable cosmic strings in the delayed scaling scenario
}

\date{\today}

\author[a.b.c]{Yifan Hu,}
\affiliation[a]{School of Fundamental Physics and Mathematical Sciences, Hangzhou Institute for Advanced Study, University of Chinese Academy of Sciences (HIAS-UCAS), 310024 Hangzhou, China}
\affiliation[b]{University of Chinese Academy of Sciences, Beijing 100049, China}
\affiliation[c]{Institute of Theoretical Physics, Chinese Academy of Sciences, Beijing 100190, China}

\author[a,d.e]{Kohei Kamada}
\affiliation[d]{International Centre for Theoretical Physics Asia-Pacific (ICTP-AP), Hangzhou/Beijing, China}
\affiliation[e]{Research Center for the Early Universe, The University of Tokyo, Bunkyo-ku, Tokyo 113-0033, Japan}

\emailAdd{huyifan23@mails.ucas.ac.cn}
\emailAdd{kohei.kamada@ucas.ac.cn} 

\abstract{Recent observations by pulsar timing arrays (PTAs) such as NANOGrav, EPTA, PPTA, and CPTA suggest the presence of nanohertz stochastic gravitational wave background (GWB). 
While such signals could be explained by gravitational waves from a network of metastable cosmic strings (CSs), standard scenarios involving the Kibble-Zurek mechanism triggered by a thermal potential face significant challenges.
Specifically, these scenarios predict a GWB spectrum inconsistent with the non-detection at higher frequencies by LIGO-Virgo-KAGRA (LVK) for CSs with relatively large string tension. 
It is also difficult to prevent the monopole forming phase transition just before the CS forming symmetry breaking, which spoils the CS network formation. 
In contrast, a delayed scaling scenario, where the CSs start to emit GWs at a later time due to the dilution during inflation, 
alleviates these issues. 
This scenario allows for a larger string tension while 
monopoles are sufficiently diluted such that the CS network safely forms. 
In this study, we clarify the spectrum of stochastic GWB from metastable CSs in the delayed scaling scenario, consistent with the PTA observations while satisfying the LVK constraints. Furthermore, we explore its potential signatures at frequencies accessible to other detectors such as LVK as well as LISA, Taiji, and TianQin or DECIGO and BBO. 
We also discuss the implications on inflation and underlying UV theories, such as the grand unified theories.}

\maketitle

\section{Introduction}

The stochastic gravitational wave background (GWB) serves as a powerful probe for exploring the early Universe cosmology. Due to their extremely weak interaction with matter, gravitational waves (GWs) preserve detailed information about their emission mechanisms and retain a record of the cosmic history. 
The ongoing and forthcoming GW observations at {\it e.g.}, LIGO~\cite{LIGOScientific:2014pky}-Virgo~\cite{VIRGO:2014yos}-KAGRA~\cite{Somiya:2011np,Aso:2013eba} (LVK) around $10^2$ Hz, LISA~\cite{LISA:2017pwj}, Taiji~\cite{Hu:2017mde,Ruan:2018tsw}, or TianQin~\cite{TianQin:2015yph,TianQin:2020hid} around mHz, and DECIGO~\cite{Seto:2001qf} or BBO~\cite{Corbin:2005ny} around 1Hz together with the observation of the polarization of the cosmic microwave background (CMB)~\cite{Planck:2018jri,LiteBIRD:2022cnt,Abazajian:2019eic} are promising avenue for understanding the early Universe cosmology. 
Recently, moreover, pulsar timing array (PTA) searches at NANOGrav~\cite{NANOGrav:2023gor}, EPTA~\cite{EPTA:2023fyk}, PPTA~\cite{Reardon:2023gzh}, and CPTA~\cite{Xu:2023wog} have reported compelling evidence for a stochastic GWB signal in the nHz frequency range. While it could be of astrophysical origin sourced by inspiraling  supermassive blackhole binaries, 
this groundbreaking discovery marks a significant milestone in the field of GW cosmology. 
Among several possible sources that could explain the PTA observations~\cite{NANOGrav:2023hvm}, cosmic strings (CSs) are outstanding, since they generate GWs with frequencies ranging from, say, $10^{-14}$ Hz to $10^{10}$ Hz~\cite{Vilenkin:1981bx,Hogan:1984is,Vachaspati:1984gt,Accetta:1988bg}, suitable for multi-band observations and offers us information on the physics beyond the Standard Model of particle physics as well as cosmic history~\cite{Vilenkin:1982hm,Hindmarsh:2011qj,Buchmuller:2013lra,Auclair:2019wcv,Gouttenoire:2019kij}. See also some recent studies that summarizes the formulation~\cite{Binetruy:2012ze,Sousa:2020sxs,Blanco-Pillado:2024aca,Schmitz:2024gds}. 

CS is a topological defect that emerges from a spontaneous symmetry breaking ${\cal G} \rightarrow {\cal H}$ where the first homotopy group is nontrivial, $\pi_1({\cal G}/{\cal H}) \not = I$~\cite{Nielsen:1973cs,Hindmarsh:1994re,Vilenkin:2000jqa}. 
Once such a spontaneous symmetry breaking occurs in the Universe, they are generated to form a network~\cite{Kibble:1976sj,Kibble:1980mv,Zurek:1985qw}. 
While long strings in the network whose length is larger than the Hubble length are stable, they continuously produce string loops.   
The distribution of the long strings is attracted to a so-called scaling regime, where the energy density of long strings is a constant fraction to the critical density~\cite{Kibble:1984hp,Albrecht:1989mk}. 
The string loops shrink and lose energy by emitting mainly Nambu-Goldstone bosons (for $G$ being a global group) or GWs (for $G$ being a local group), such that CSs never dominate the energy density of the Universe. 
The string loops are continuously generated with sizes proportional to the Hubble length, resulting in GWs from local CSs spanning a broad range of frequency bands. 

However, GWs from ``metastable'' CSs are consistent with them~\cite{NANOGrav:2023hvm}. 
CSs can be metastable if there is another symmetry breaking that breaks down to ${\cal G}$, ${\cal F}\rightarrow {\cal G}$, and the second homotopy group is non-trivial, $\pi_2({\cal F}/{\cal G}) \not = I$ while the total homotopy group ${\cal F}/{\cal H}$ is trivial. 
Note that such a symmetry breaking chain is ubiquitous in the grand unified theories (GUTs)~\cite{Jeannerot:2003qv}. 
If these symmetry breaking scales are hierarchical, CSs can generated in cosmic history to form a network once but eventually decay~\cite{Leblond:2009fq,Buchmuller:2019gfy,Buchmuller:2020lbh,Buchmuller:2021mbb,Buchmuller:2023aus,Chitose:2024pmz} through the monopole-antimonopole pair creation~\cite{Vilenkin:1982hm,Preskill:1992ck,Chitose:2023dam}. 
CSs with string tension around $10^{-8} < G\mu < 10^{-5}$ with $G$ being the gravitational constant and the decay time around redshift $z\sim 10^{10}$  offer a good fit to the PTA measurements, 
which allows us to examine both of two symmetry breaking scales. 

However, if the symmetry breaking is triggered by the Kibble-Zurek mechanism with a thermal potential~\cite{Kibble:1976sj,Kibble:1980mv,Zurek:1985qw}, GWs from CSs with a slightly large string tension, $G\mu>2 \times 10^{-7}$, should have already been detected by LVK observations~\cite{KAGRA:2021kbb}, unless the reheating temperature is relatively low~\cite{Chitose:2024pmz}.
This constraint can be also relaxed if we consider the delayed scaling scenario, where the symmetry breaking takes place during inflation and string loop production and hence GW emissions are delayed due to the dilution of the long strings~\cite{Lazarides:1984pq,Shafi:1984tt,Vishniac:1986sk,Kofman:1986wm,Yokoyama:1988zza,Yokoyama:1989pa,Nagasawa:1991zr,Basu:1993rf,Freese:1995vp,Kamada:2012ag,Linde:2013aya,Kamada:2014qta,Zhang:2015bga,Ringeval:2015ywa,Guedes:2018afo,Cui:2019kkd}. 
If future observations detect the feature of the GWs from CSs with the delayed scaling scenario, another information of early Universe cosmology, inflation, can be obtained.

In this article, we examine carefully the GW spectrum from metastable CSs in the delayed scaling scenario. 
We identify the characteristic frequencies of the spectral breaks and spectral indices of each part, which are consistent with previous studies. 
We find that to explain the PTA measurements a larger string tension, $G\mu>10^{-6}$, is allowed while satisfying the LVK constraints, if the scaling starts at a later time, $z<10^{14}$. 
The GW spectrum shows a plateau at the intermediate frequencies, typically, $10^{-6} \mathrm{Hz} \lesssim f \lesssim 10^{-2} \mathrm{Hz}$, with a spectral break to $f^{-1/3}$ decay, which can be detected by the future GW detectors, such as LISA, Taiji, TianQin, DECIGO, or BBO. 
They will determine the tension as well as the times of the onset of scaling evolution and string decay. 
The string tension takes the maximum value, $G\mu\sim 3 \times 10^{-5}$, 
which is determined by the condition that the cutoff of the spectrum comes at a frequency lower than those for PTA measurements. 
It has been already pointed out by the NANOGrav collaboration, 
but we conclude that it does not change in the delayed scaling scenario. 
We also discuss the implication on inflation, by showing a working example of the inflationary scenario. 

The remainder of this article is structured as follows.
In the next section,  we review the scaling evolution of CS networks, including how this behavior is modified for metastable CSs. We also discuss the incorporation of the delayed scaling scenario. 
Section~\ref{sec:GWspectrum} presents a numerical evaluation of the GW spectrum, complemented by an analytical estimate to ensure consistency. The high-frequency behavior of the spectrum and the observational prospects are also discussed.  
In Sec.~\ref{sec:implication}, we explore the implementation of this scenario within a concrete model of the inflationary Universe, highlighting the essential features required for its realization.  
Finally, Sec.~\ref{sec:summary} summarizes our findings.

\section{Metastable cosmic strings and delayed scaling scenario}
\subsection{Cosmic string network}

We start with the review of the time evolution of the network of stable CSs.
After an initial transient period subsequent to the spontaneous symmetry breaking, 
the network enters a ``scaling regime'', where the ratio between the 
energy density of CSs with the superhroizon length and the critical density is constant. 
In the case of local CSs, oscillating loops lose their energy by emitting GWs and shrink. 
Such a picture is strongly supported by both analytic models~\cite{Kibble:1984hp,Albrecht:1989mk,Caldwell:1991jj,Martins:1996jp,Martins:2000cs,Sousa:2013aaa,Sousa:2014gka,Sousa:2020sxs} and numerical simulations~\cite{Vilenkin:1981bx,Vachaspati:1984gt,Accetta:1988bg,Martins:2005es,Ringeval:2005kr,Lorenz:2010sm,Blanco-Pillado:2011egf,Sanidas:2012ee,Binetruy:2012ze,Blanco-Pillado:2013qja,Blanco-Pillado:2017oxo,Blanco-Pillado:2017rnf}. 
In the following, we model the evolution of CS network based on Ref.~\cite{Blanco-Pillado:2011egf,Blanco-Pillado:2013qja} (Blanco-Pillado-Olum-Shlaer (BOS) model) for concreteness, 
particularly focusing on the number density of the string loops.

The CS is characterized by the tension, $\mu$, which is related to the symmetry breaking scale $v_\mathrm{cs}$ as $\mu \sim  \pi v_\mathrm{cs}^2$~\cite{Vilenkin:2000jqa}.
In the scaling regime, the energy of long strings that enter the horizon is balanced by loop formation such that the loop production rate is proportional to $t^{-4}$ during radiation dominated era with $t$ being the physical time. 
We suppose that the loops are generated with a constant length proportional to the horizon length, $l = \alpha t$, where $\alpha$ is a constant that determines the ratio between the loop length and Hubble length. 
Then the loops shrink by emitting GWs with the total power $\Gamma G \mu$
with $\Gamma$ being a dimensionless constant. 
Hereafter we take $\alpha=0.1$ and $\Gamma =50$  according to the numerical simulation~\cite{Blanco-Pillado:2013qja,Blanco-Pillado:2017oxo}. 
Noting that the energy of a string loop with a length $l$ is given by $\mu l$, 
the CS length $l(t)$ at time $t$, which had a length $l'$ at time $t'$, is expressed as
\begin{equation}
    l(t) = l' - \Gamma G \mu (t-t'). 
\end{equation}
Taking into account the dilution due to cosmic expansion after the generation of the loop, the number density of loops with a length $l$ at time $t$ is modeled as
\begin{equation}
    n^\mathrm{r} (l,t)= \frac{0.18}{t^{3/2} (l+\Gamma \mu t)^{5/2}} \Theta(\alpha t-l)\Theta(t_\mathrm{eq}-t),
    \label{n^r}
\end{equation}
in the radiation dominated era, where $t_\mathrm{eq}$ is the time at the matter radiation equality, and $\Theta(x)$ is the step function. 
In the matter dominated era, loops produced during radiation dominated era and those produced during matter dominated era have experienced different dilution history. Thus we distinguish them to write
\begin{equation}
    n^\mathrm{rm} (l,t)= \frac{0.18 (2 H_0 \Omega_\mathrm{r}^{1/2})^{3/2}}{(l+\Gamma G \mu t)^{5/2}} (1+z(t))^3 \Theta(\alpha t_\mathrm{eq} -{\bar l}(t_\mathrm{eq};t,l) ) \Theta(t-t_\mathrm{eq}),
\end{equation}
for the former where $\Omega_\mathrm{r}$ is the density parameter of radiation today and we define ${\bar l}({\bar t};t,l) \equiv l - \Gamma G \mu ({\bar t} - t)$, while 
\begin{equation}
    n^\mathrm{m} (l,t)= \frac{0.27 - 0.45 (l/t)^{0.31}}{t^2 (l+\Gamma G \mu t)^2} \Theta(\alpha t-l) \Theta(t-t_\mathrm{eq}) ,
\end{equation}
for the latter.

\subsection{Metastable cosmic strings}

Once we assume a richer structure in the symmetry breaking pattern, we can consider the case where the CSs decay by the monopole-antimonopole pair creation on the string through quantum tunneling, 
which is often the case in the models of grand unified theories (GUTs)~\cite{Jeannerot:2003qv}. 
Here we model this scenario following Refs.~\cite{Vilenkin:1982hm,Preskill:1992ck,Leblond:2009fq,Buchmuller:2021mbb}. 

The decay rate per string unit length is expressed by
\begin{equation}
    \Gamma_d = \frac{\mu}{2\pi} \exp[-\pi \kappa], 
\end{equation}
where $\kappa$ is determined by the hierarchy between two symmetry breaking scales.
The decay rate determines the time when almost all the long strings experiences the monopole-antimonopole pair creation, $t_s \equiv 1/\Gamma_d^{1/2}$. 
For $\sqrt{\kappa} \gg 1$ when the hierarchy between two symmetry breaking scales is large enough, 
the decay rate is sufficiently law such that the string decay does not occur much until today. 
If the hierarchy between them is relatively small, $\sqrt{\kappa} \sim {\cal O}(1)$, the string decay occurs in the early Universe such that the GW signal can be affected. 
In the following we consider the case $\sqrt{\kappa}\sim 7.5-8$, in which case string decays during radiation dominated era and GW signal around the PTA frequencies are modified but not significantly damped~\cite{Buchmuller:2020lbh}
to explain the PTA measurements~\cite{NANOGrav:2023hvm}.
Indeed, such values of $\kappa$ can be obtained in typical GUT models~\cite{King:2021gmj,King:2023wkm}. 

While at $t<t_s$, loops behave quite similar to stable string loops,
but with a suppression factor due to partial decay through the monopole-antimonopole production on the loop which results in the production of short subhorizon segments.
Consequently, we estimate the loop number density in this period as
\begin{equation}
     \mathring{n}^\text r_<(l,t)=\frac{0.18}{t^{3/2}(l+\Gamma G\mu t)^{5/2}}e^{-\Gamma_d[l(t-l/\alpha)+\Gamma G\mu(t-l/\alpha)^2/2]}\Theta(\alpha t-l)\Theta(t_{\text{eq}}-t). \label{LoopNumberRR<}
\end{equation}
At $t>t_s$, loops are no longer produced from long strings, 
and those produced at $t>t_s$ are simply diluted. 
Taking into account the length of the loop at $t_s$, the number density of loops is evaluated as
\begin{equation}
    \mathring{n}^\text r_>(l,t)=\frac{0.18}{t^{3/2}(l+\Gamma G\mu t)^{5/2}}e^{-\Gamma_d[l(t-t_s)+\Gamma G\mu(t-t_s)^2/2]}\Theta\left(\alpha t_s-{\bar l}(t_s;t,l)\right)\Theta(t_{\text{eq}}-t). 
    \label{LoopNumberRR>}
\end{equation}
At $t<t_\mathrm{eq}$, loops still may survive.
In a similar way to the case of stable string, we express the loop number density at this epoch as
\begin{equation}
    \mathring{n}^\mathrm{rm}_>(l,t) = \frac{0.18 (2 H_0 \Omega_\mathrm{r}^{1/2})^{3/2}}{(l+\Gamma G \mu t)^{5/2}} (1+z(t))^3 e^{-\Gamma_d[l(t-t_s)+\Gamma G\mu(t-t_s)^2/2]}\Theta\left(\alpha t_s-{\bar l}(t_s;t,l)\right) \Theta(t-t_\mathrm{eq}). 
    \label{LoopNumberRM>}
\end{equation}

\subsection{Delayed scaling scenario}

Thus far we have not specified when the CS network is formed and enters the scaling regime. 
In the case of the Kibble-Zurek mechanism driven by the thermal potential, the phase transition takes place roughly around $T \sim v_\mathrm{cs}\sim \sqrt{\mu}$ with $T$ being the temperature of the Universe. 
If the temperature is high enough, the thermal friction suppresses the motion of long strings and prevents them from entering the scaling regime. 
The thermal friction becomes negligible roughly at $T\sim G^{1/2} \mu$, and the CS network would enter the scaling regime~\cite{Vilenkin:1991zk,Vilenkin:2000jqa,Gouttenoire:2019kij}. 
Such a scenario has problems in explaining the PTA measurements. 
First, CSs with tension $10^{-7}<G\mu<10^{-5}$ also predict stochastic GWB around $10^{2}$ Hz, which should have been detected by LVK measurements. 
Indeed, LVK has given merely the upper bound on the string tension at 25 Hz as~\cite{KAGRA:2021kbb}
\begin{equation}
    \Omega_\mathrm{GW} \leq 1.7 \times 10^{-8}, \quad \text{corresponding to} \quad G\mu \leq 2 \times 10^{-7}. \label{LVKconst}
\end{equation}
Second, since the symmetry breaking scales for strings and monopoles are relatively close~\cite{Preskill:1992ck,Chitose:2023dam}, the symmetry breaking with monopole production would also have occurred in the thermal history of the Universe. 
In this case, the CSs are formed in a way that to connect monopoles and antimonopoles distributed at subhorizon and disappear quickly without forming a CS network.

These problems can be avoided if symmetry breaking occurs during inflation~\cite{Lazarides:1984pq,Shafi:1984tt,Vishniac:1986sk,Kofman:1986wm,Yokoyama:1989pa,Yokoyama:1988zza,Nagasawa:1991zr,Basu:1993rf,Freese:1995vp,Kamada:2012ag,Linde:2013aya,Kamada:2014qta,Zhang:2015bga,Ringeval:2015ywa,Guedes:2018afo,Cui:2019kkd}. 
The CSs are diluted during inflation such that distance between long strings becomes subhorizon to enter the scaling regime at a later time sufficiently after inflation and reheating. 
The monopoles can be also diluted to sufficiently superhorizon scales before the CS forming symmetry breaking and hence the CS network can form.
We will discuss more quantitative implication of inflation and symmetry breaking scales in Sec.~\ref{sec:implication}.

The separation of the long strings, $L$, is initially order of the Hubble length during inflation, while depending on some of the detail of the potential of the symmetry breaking field~\cite{Vishniac:1986sk,Yokoyama:1989pa,Nagasawa:1991zr,Kamada:2011bt,Kamada:2012ag}, and increases due to the exponential expansion of the Universe, $\exp[\Delta N_\mathrm{cs}]$, where $\Delta N_\mathrm{cs}$ is the number of e-folds between the symmetry breaking and inflation end. 
Consequently,  the ratio between the string separation and the Hubble length, $HL$, becomes much larger than the unity, $HL \gg 1$. 
After inflation, it decreases because of the difference in the growth rate of the scale factor and Hubble length. 
Once it becomes order of the unity, $HL \sim 1$, long strings start producing 
loops, and eventually enters the scaling regime, 
which takes a few more e-folding numbers~\cite{Kamada:2014qta,Ringeval:2015ywa,Cui:2019kkd}. 
One may use the velocity-dependent-one-scale (VOS) model~\cite{Kibble:1984hp,Martins:1996jp,Martins:2000cs} 
to determine the time evolution of the separation of the long strings 
and consequently the loop number density. 
Instead, in the following we model the system that the loop number density is given by Eq.~\eqref{LoopNumberRR<} from a time $t_\mathrm{sc}$ and evaluate the contribution on GWB at $t>t_\mathrm{sc}$. 
Note that we require $t_\mathrm{sc}<t_s$, otherwise CSs decay before entering the scaling regime and do not produce loops.

\section{Gravitational wave spectrum} \label{sec:GWspectrum}
With the properties of the evolution of the CS network discussed in the previous section, we now examine the GW spectrum from the metastable CSs in the delayed scaling scenario. 
While several approaches have been studied in evaluating the GW spectrum
in both analytic and numerical methods, 
here we adopt the method developed in Ref.~\cite{Vachaspati:1984gt,Blanco-Pillado:2017oxo,Blanco-Pillado:2017rnf} for concreteness.
We first review the GW spectrum from the metastable CSs 
and extend the formula to the delayed scaling scenario. We numerically evaluate the
GW spectrum 
and provide a fitting formula supported by analytical insights. 

In this study, we quantify the GWB 
in terms of the energy density of gravitational waves per unit logarithmic frequency interval $\Omega_\mathrm{gw}$, normalized to the critical density at the present time $t_0$ as 
\begin{equation}
    \Omega_{\text{gw}}(t_0,f)=\frac{1}{\rho_\text{c}}\frac{\text d ~\rho_{\text{gw}}(t_0,f)}{\text d~\text{ln}f},
    \label{Omega_gw}
\end{equation}
where critical energy density is denoted by $\rho_{\text{c}}=3H_0^2/8\pi G$ 
with $H_0$ being the present Hubble parameter, and
$\rho_\mathrm{gw}$ is the energy density of the gravitational waves. 
The stochastic GWs from CSs are understood as the summation of those from CS loops that are continuously produced during the scaling evolution.
Given a present GW frequency, we add up the GWs from all the loops throughout the cosmic history that contribute to that frequency.
 $\text{d}\rho_{\text{gw}}/\text{d}f$ can be then expressed as
\begin{equation}
    \frac{\text d\rho_{\text{gw}}(t_0,f)}{\text d \ln f}=G\mu^2 f\sum_{k=1}^{k_\mathrm{max}} C_k(t_0,f)p_k,
    \label{energy density}
\end{equation}
where $\mu$ is string tension. 
In the following, we describe the meaning of each terms in this expression.

In our treatment, we sum up the contributions from each oscillation mode of the string loops with the length $l$, such that the GW frequency at the emission is estimated as $f'=2k/l$
with $k$ indicating the
$k$-th harmonic mode.\footnote{We do not take into account the contributions from string segments as well as long strings, which are often subdominant. Note that string segments would not contribute to GWB if monopoles carry unconfined flux (which is the case for many GUT models) since they lose energy mainly by emitting massless vector bosons.} 
The upper bound of the harmonic mode $k_\mathrm{max}$ should be taken to be infinity, as long as infinitely thin Nambu-Goto (NG) strings are concerned. 
We will discuss the treatment of $k_\mathrm{max}$ in evaluating the GWB later.
$p_k$ is the GW power radiated by the $k$-th harmonic mode of a loop in units of  $G \mu^2$.
It has been observed that the power obeys the scaling law with respect to $k$ as
\begin{equation}
    p_k=\frac{\Gamma}{\zeta (q)}k^{-q},
    \label{Single loop power spectra}
\end{equation}
where $\zeta(x)$ is Riemann zeta function (infinitely thin strings are supposed) such that the total GW emission power is $P=\Gamma G \mu^2$. 
Index $q$ depends on the origin of the GW and 
is $4/3$, $5/3$ and $2$ corresponding cusps, kinks and kink-kink collisions respectively~\cite{Vachaspati:1984gt,Damour:2001bk,Binetruy:2009vt}. 
Hereafter we take $q=4/3$, which is also consistent with the results of numerical simulations~\cite{Blanco-Pillado:2017oxo}.
Another coefficient in Eq.\eqref{energy density}, $C_k(t_0,f)$ 
counts the number of string loops whose $k$-th modes contribute to the GW spectrum at a frequency $f$ today throughout the cosmic history,
given by
\begin{equation}
    C_k(t_0,f)=\frac{2k}{f^2}\int_{0}^\infty\frac{\text dz'}{H(z')(1+z')^6}n\left(\frac{2k}{(1+z')f},t(z')\right),
    \label{C_k}
\end{equation}
where $n(l,t)$ is density of non-self-interacting, sub-horizon string loops with proper length $l$ at cosmic time $t$, given in the previous section. 
We here take into account the dilution of the number density as well as the GW frequency due to the cosmic expansion.

\begin{figure}[H]
    \centering
    \includegraphics[width=0.8\linewidth]{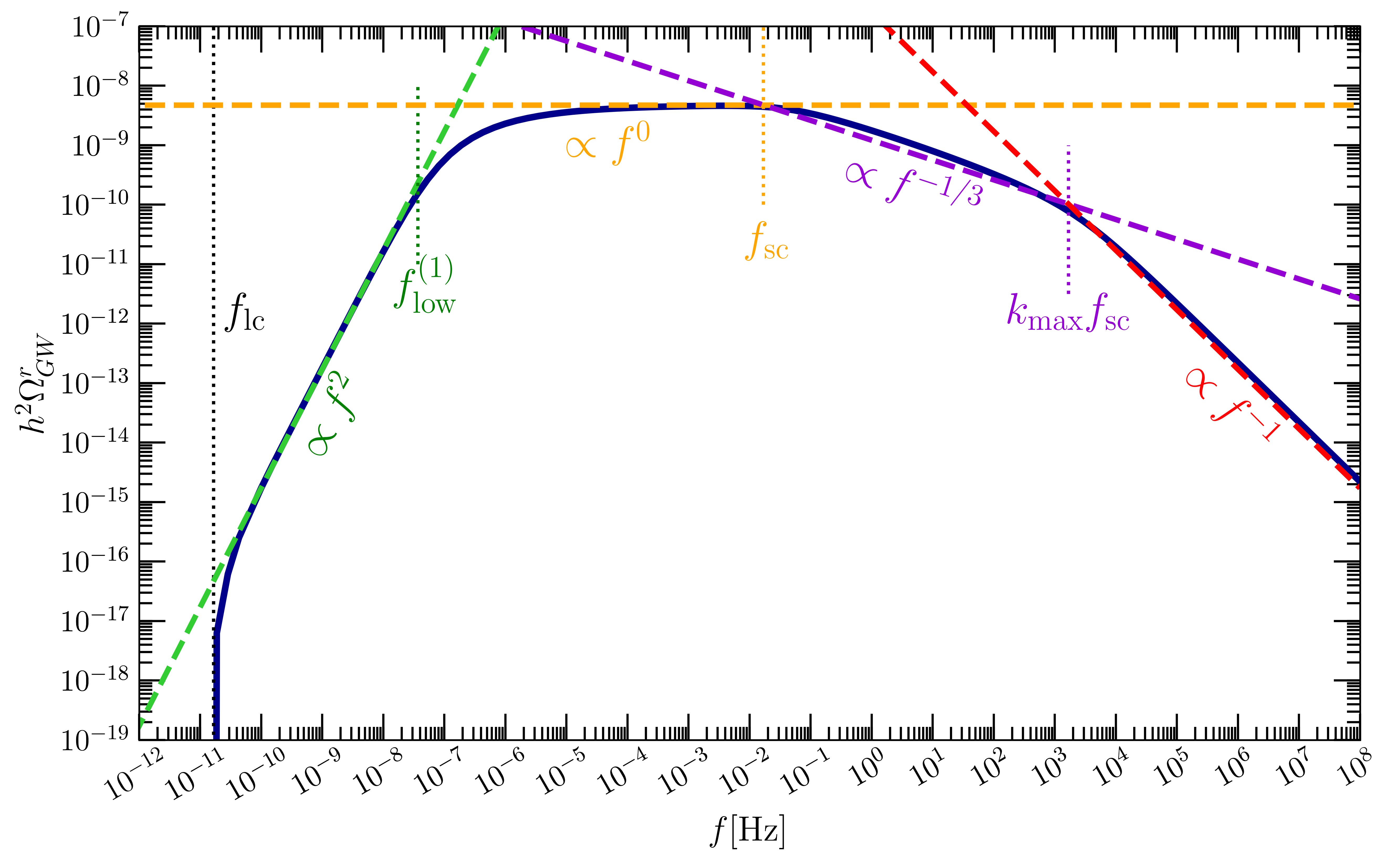}
    \caption{The typical GW spectrum from metastable CS network in the delayed scaling scenario is shown. The parameters are chosen as $G\mu={10^{-8}}, \sqrt{\kappa} = {8}$,  $z_\mathrm{sc}= {10^{11}}$, and $k_\mathrm{max}={10^5}$. We see a low frequency cutoff, an increase proportional to $f^2$, a plateau at an intermediate frequency range, and a decay proportional to $f^{-1/3}$ followed by the one proportional to $f^{-1}$. The spectrum is well fitted by Eqs.~\eqref{omegagwlow}, \eqref{omegagwplateau}, \eqref{omegagwhigh13}, and \eqref{omegagwhigh1}, with the typical frequencies, $f_\mathrm{lc}$~\eqref{flc}, $f_\mathrm{low}^{(1)}$~\eqref{zm}, $f_\mathrm{sc}$, and $k_\mathrm{max} f_\mathrm{sc}$~\eqref{fhigh}. }
    \label{TypicalSpectrum}
\end{figure}

With these formulation, we now evaluate the GW spectrum 
numerically integrating Eqs.\eqref{Omega_gw}-\eqref{C_k} together with the loop number density (Eqs.~\eqref{LoopNumberRR<} and \eqref{LoopNumberRR>}).
Note that we consider the case with $t_s > t_\mathrm{eq}$ as we mentioned. In this case, $\mathring{n}^\mathrm{rm}_>(l,t)$ is exponentially suppressed, and hence we do not take into account it. 
We then evaluate the $z'$ integration in Eq.~\eqref{C_k} from $z_\mathrm{eq}$ to $z_\mathrm{sc}$, where $z_\mathrm{eq}$ and $z_\mathrm{sc}$ are the redshift at matter-radiation equality and the time when the CS network enters the scaling regime, respectively.
For the background expansion history of the Universe, we use the Hubble parameter for the radiation dominated era,
\begin{equation}
    H(z)=(1+z)^2H_r \quad \text{and} \quad t(z)=\frac{1}{2(1+z)^2H_r}, \quad \text{where} \quad H_r=H_0\sqrt{\Omega_r}, 
    \label{Cosmological parameters}
\end{equation}
with $\Omega_r={9.1476}\times10^{-5}$ being the radiation energy fraction today~\cite{2009ApJ...707..916F,Planck:2018vyg}, and $H_0=100h~\text{km/s/Mpc}$  being the Hubble constant today with $h=0.674$~\cite{Planck:2018vyg}. Here we do not take into account the change of the effective number of degree of freedom. 
A typical GW spectrum is shown in Fig.~\ref{TypicalSpectrum}. 

The feature of the GW spectrum can be understood as follows.
Substituting Eqs. \eqref{LoopNumberRR<}, \eqref{LoopNumberRR>}, and \eqref{Cosmological parameters} into Eq.~\eqref{C_k}, 
the coefficient functions are given by
\begin{align}
    C_k(t_0,f)=&
    \frac{0.18\times32 H_r^3k}{f^2}\Bigg[\int_{z_\text{eq}}^{z_\text s}\dd z\frac{\text e^{-\Gamma_d\left[\frac{k}{fH_r(1+z)}\left(\frac{1}{(1+z)^2}-\frac{1}{(1+z_\text{s})^2}\right)+\frac{\Gamma G\mu}{8H_r^2}\left(\frac{1}{(1+z)^2}-\frac{1}{(1+z_\text{s})}\right)^2\right]}}{\left[\frac{4kH_r}{f}(1+z)+\Gamma G\mu\right]^\frac{5}{2}} \notag \\
    & \times \Theta \left(\alpha t_s - {\bar l}\left(t_s;t(z'),\frac{2k}{(1+z')f}\right) \right) \notag 
    \\
    +&\int_{z_\text s}^{z_\text{sc}}\text dz\frac{\text e^{-\frac{\Gamma_d}{(1+z)^2}\big[\frac{2k}{f}(\frac{1}{2(1+z)H_r}-\frac{2}{\alpha f})-\Gamma G\mu(\frac{1}{2(1+z)H_r}-\frac{2k}{\alpha f})^2\big]}}{\left[\frac{4kH_r}{f}(1+z)+\Gamma G\mu\right]^{\frac{5}{2}}} \Theta \left( \alpha t(z')-\frac{2k}{(1+z')f}\right) \Bigg]. 
    \label{C_k(t_0,f)}
\end{align}
Note that we require 
$t_{\text{sc}}<t_\text s$ or equivalent $z_{\text{sc}}>z_\text s$. 
First, the step function function in Eq.~\eqref{C_k(t_0,f)} gives the lower bound of the frequency as 
\begin{equation} 
    f>f_\mathrm{lc} = \frac{4 (1+z_s)}{\alpha}H_r,  \label{flc}
\end{equation}
which explains the cutoff at the lower frequency.

We can now omit the step function in Eq.~\eqref{C_k(t_0,f)} as long as $f>f_\mathrm{lc}$. 
We then approximate the exponential suppression factor as
$\Theta(z-z_m)$ with 
\begin{equation}
   z_m \equiv \left\{ \begin{array}{ll} 
   \left(\dfrac{4 z_s k H_r}{f}\right)^{1/3} z_s \equiv z_f& \text{for} \quad f_\mathrm{lc} < f < f_\mathrm{low}^{(k)} \equiv 4 z_s k H_r \left(\dfrac{\Gamma G \mu}{2} \right)^{-3/4} \\
   \\
   \left(\dfrac{\Gamma G \mu}{2} \right)^{1/4} z_s  \equiv z_e& \text{for} \quad f> f_\mathrm{low}^{(k)}
   \end{array}\right. , \label{zm}
\end{equation}
 such that we have a following expression as
\begin{equation}
    C_k(t_0,f) \simeq \frac{0.18\times16H_r^2}{3f}\left[\left(\frac{4kH_r}{f}(1+z_m)+\Gamma G\mu\right)^{-\frac{3}{2}}-\left(\frac{4kH_r}{f}(1+z_{\text{sc}})+\Gamma G\mu\right)^{-\frac{3}{2}}\right].
    \label{C_k(t_0,f) for metastable string}
\end{equation}
At low frequency, $f_\mathrm{lc} < f < f_\mathrm{low}^{(1)}$,   the fundamental mode ($k=1$) in the first term, $4 H_r (1+z_m)/f$ in the parentheses dominates the GW spectrum with $C_1(t_0,f)$ being evaluated as 
\begin{equation}
    C_1(t_0,f) \simeq  \frac{0.18\times16H_r^2}{3f} \left(\frac{4 H_r}{f} z_f \right)^{-3/2}. 
\end{equation}
Consequently, the GW spectrum is approximated as
\begin{align}
    \Omega_\mathrm{GW}h^2 &\simeq  \frac{0.18 \times 8 \pi G^2 \mu^2}{9 H_0^2/h^2} \frac{\Gamma}{\zeta (4/3)}   z_s^{-2} f^2\notag \\
    & \simeq 6.6 \times 10^{-11} \left(\frac{G \mu}{10^{-5}}\right)^2 \left(\frac{\Gamma}{50}\right) \left(\frac{z_s}{10^{10}}\right)^{-2} \left(\frac{f}{10^{-8} \mathrm{Hz}}\right)^2 \quad \text{for} \quad f_\mathrm{lc} < f < f_\mathrm{low}^{(1)} . \label{omegagwlow}
\end{align}
At higher frequency $f>f_\mathrm{low}^{(k)}$, $\Gamma G \mu$ in the first term in Eq.~\eqref{C_k(t_0,f) for metastable string} dominates such that it is approximated as
\begin{equation}
    C_k(t_0,f) \simeq \frac{0.18\times16H_r^2}{3f} (\Gamma G \mu)^{-3/2}. 
\end{equation}
The GW spectrum shows a plateau as
\begin{align}
    \Omega_\mathrm{GW}h^2 &\simeq \frac{0.18 \times 128 \pi (G \mu)^{1/2} H_r^2}{9 \Gamma^{3/2} H_0^2/h^2} \sum_k p_k  \notag \\
    & \simeq 1.5 \times 10^{-7}  \left(\frac{\Gamma}{50}\right)^{-1/2} \left( \frac{G\mu}{10^{-5}}\right)^{1/2} \left(\frac{h}{0.68}\right)^2. \label{omegagwplateau}
\end{align}
Note that the plateau has contributions from all the harmonic mode, the GW spectrum reaches at the plateau at the frequency, $f_\mathrm{low}^{({\bar k})}={\bar k} f_\mathrm{low}^{(1)}$ with ${\bar k} \simeq 10^{2-3}$, when $\sum_k^{\bar k} k^{-4/3}$ gets sufficiently close to $\zeta(4/3)$. 
This low frequency behavior coincides with the GW spectrum in the metastable CSs~\cite{Buchmuller:2021mbb}, independent of the way how the system starts to emit GWs, since the string loops formed at earlier times do not contribute to low frequency spectra. 

At much higher frequencies, 
\begin{equation}
    f>f_\mathrm{high}^{(k)} \equiv \frac{4k H_r}{\Gamma G \mu} z_\mathrm{sc}, \label{fhigh}
\end{equation}
the second term in Eq.~\eqref{C_k(t_0,f) for metastable string} is no longer negligible, such that it is approximated as 
\begin{equation}
    C_k(t_0,f) \simeq \frac{0.18\times 32H_r^3 k}{f^2} (z_\mathrm{sc}-z_e) (\Gamma G \mu)^{-5/2}, \label{ckhigh}
\end{equation}
which shows a weak dependence on $z_e$, but negligibly small for $z_\mathrm{sc} \gg z_e$. 
With an artificially introduced upper bound of the harmonic mode, $k_\mathrm{max}$, at $f_\mathrm{high}^{(1)} \equiv f_\mathrm{sc} < f < f_\mathrm{high}^{(k_\mathrm{max})} = k_\mathrm{max} f_\mathrm{sc} $, the GW spectrum can be approximated as
\begin{align}
    \Omega_\mathrm{GW} h^2 & \simeq \frac{0.18 \times 128 \pi  H_r^2}{3 \Gamma^{1/2} \zeta(4/3) H_0^2/h^2} (G\mu)^{1/2} \left(\sum_{k=1}^{\lfloor f/f_\mathrm{sc} \rfloor} \frac{  f_\mathrm{sc}}{2f} k^{-1/3}+ \sum_{k=\lceil f/f_\mathrm{sc} \rceil}^{k_\mathrm{max}} \frac{k^{-4/3}}{3} \right) \notag \\
    &\simeq \frac{0.18 \times 64 \pi  H_r^2}{3 \Gamma^{1/2} \zeta(4/3) H_0^2/h^2}(G\mu)^{1/2} \left(\frac{f}{f_\mathrm{sc}} \right)^{-1/3} \notag \\
    & \simeq 1.5 \times 10^{-7}  \left(\frac{\Gamma}{50}\right)^{-1/2} \left( \frac{G\mu}{10^{-5}}\right)^{1/2} \left(\frac{h}{0.674}\right)^2 \left(\frac{f}{f_\mathrm{sc}}\right)^{-1/3}, \label{omegagwhigh13}
\end{align}
where we have assumed that $k_\mathrm{max} \gg 1$. See Ref.~\cite{Buchmuller:2021mbb,Schmitz:2024gds} for the details of the approximation. 
Finally, at $f>k_\mathrm{max} f_\mathrm{sc}$, we have $k_\mathrm{max}<f/f_\mathrm{sc}$ so that the second term in Eq.~\eqref{omegagwhigh13} does not give contributions. Consequently, we obtain
\begin{align}
    \Omega_\mathrm{GW} h^2 & \sim \frac{0.18 \times 64 \pi  H_r^2}{3 \Gamma^{1/2} \zeta(4/3) H_0^2/h^2} (G\mu)^{1/2} k_\mathrm{max}^{-1/3} \left(\frac{f}{k_\mathrm{max} f_\mathrm{sc}}\right)^{-1} \notag \\
    &\sim 1.5 \times 10^{-7}  \left(\frac{\Gamma}{50}\right)^{-1/2} \left( \frac{G\mu}{10^{-5}}\right)^{1/2} \left(\frac{h}{0.674}\right)^2 k_\mathrm{max}^{-1/3}\left(\frac{f}{k_\mathrm{max} f_\mathrm{sc}}\right)^{-1}. \label{omegagwhigh1}
\end{align}
This behavior is the same to the high frequency behavior of the GWs from stable CS network. 
While at high frequencies GWs contain contributions from string loops formed at later times, they are negligibly small as long as $z_\mathrm{sc} \gg z_e$, see Eq.~\eqref{ckhigh}. 
Recently it is pointed out that there can be a cutoff at high frequency for stable CSs with smaller tension~\cite{Schmitz:2024hxw,Schmitz:2024gds}. 
In principle its appearance would differ in metastable CSs, but with the amplitude of the CS tension of our present interest, we do not have the cutoff at high frequency. 
These analytic formula fit the result of our numerical calculation, see Fig.~\ref{TypicalSpectrum}.

In summary, with the numerical calculation as well as the analytic estimates in the above, we conclude that the GW spectrum from the metastable CSs in the delayed scaling scenario is given as follows:
The spectrum grows as $f^{2}$ from $f=f_\mathrm{lc}$ to $f_\mathrm{low}^{(1)}$. At $f\sim f_\mathrm{low}^{(1)}$,  the spectrum reaches a plateau. 
This plateau 
continues up to $f_\mathrm{sc}$, where the contribution around $z\simeq z_\mathrm{sc}$
in \eqref{C_k(t_0,f) for metastable string} is no longer negligible.
At high frequency range, the spectrum decreases as $1/f^{1/3}$ at $f_{\text {sc}}<f<k_{\text{max}}f_{\text{sc}}$ and falls off as $1/f$ at  $f>k_{\text{max}}f_{\text{sc}}$.

\subsection{On the spectral break at high frequency}

\begin{figure}[H]
    \centering
    \includegraphics[width=0.8\linewidth]{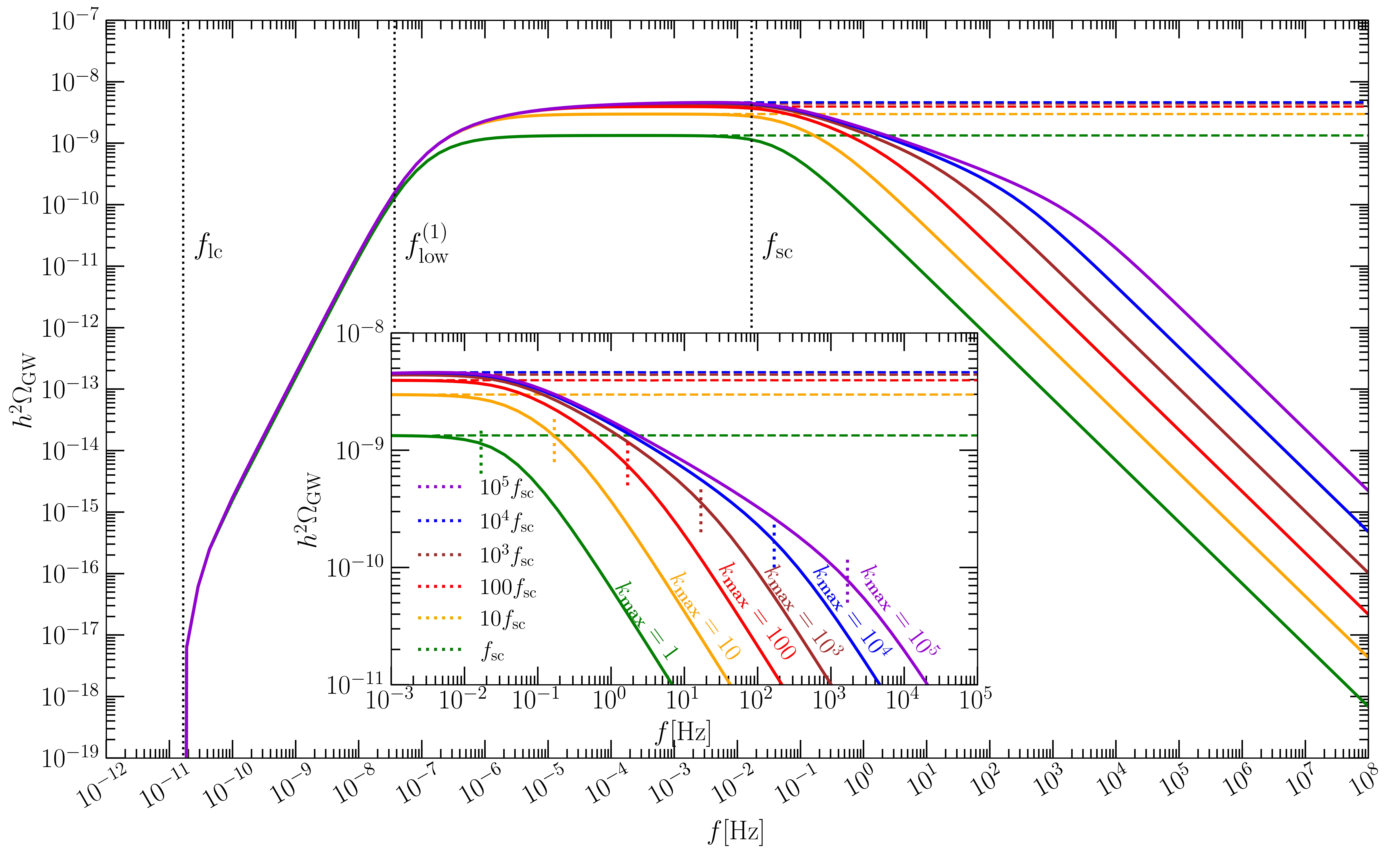}
    \caption{The GW spectrum from metastable CS network in the delayed scaling scenario with several choices of $k_\mathrm{max}$ is shown. The parameters are chosen as $G\mu=10^{-8}, \sqrt{\kappa} = 8$, and $z_\mathrm{sc}= 10^{11}$. Green, yellow, red, brown, blue, and purple solid lines correspond to the cases with $k_\mathrm{max}=1, 10, 10^2, 10^3, 10^4, 10^5$, respectively, while dashed lines correspond to the cases with $z_\mathrm{sc} = {10^{32}}$, where $f_\mathrm{sc} \gg 10^8$ Hz,  with the same color choices. The low frequency tail and plateau converges for sufficiently large $k_\mathrm{max} \gtrsim 10^2$, but high frequency tail is sensitive to the choice of $k_\mathrm{max}$. We see the spectral break  at the frequency $f=k_\mathrm{max} f_\mathrm{sc}$, as explained in Eq.~\eqref{omegagwhigh1}.} 
    \label{Spectrum with different harmonic mode}
\end{figure}

In the discussion above, we have introduced an artificial parameter, $k_\mathrm{max}$. 
For infinitely thin NG strings, $k_\mathrm{max}$ should be infinite. 
The introduction of the upper bound of the number of harmonic mode is purely practical; we cannot sum up the contributions of infinite number of harmonic modes numerically. 
While the GW spectrum at low frequency converges at sufficiently large $k_\mathrm{max}$, the high frequency tail is sensitive to $k_\mathrm{max}$. Figure~\ref{Spectrum with different harmonic mode} shows the dependence of the GW spectrum on $k_\mathrm{max}$. 
One can see that Eqs.~\eqref{omegagwhigh13} and \eqref{omegagwhigh1} work well and  $f^{-1/3}$ tail at high frequency is sensitive to the choice of $k_\mathrm{max}$. 
If the NG description is perfectly correct, we would not have the break from $f^{-1/3}$ to 
$f^{-1}$ decay and $f^{-1/3}$ decay would continue to infinite. 
Then, what is the realistic behavior of the GW spectrum at high frequency?
We discuss it in this subsection. 
Note that similar discussions can also be found in Ref.~\cite{Schmitz:2024gds}. 

Even though string is viewed as an infinitely thin line in the NG approximation, it has a non-zero thickness. 
It is evaluated as $\delta\sim(\mu/2\pi)^{-1/2}$, which depends only on the vacuum expectation value of the symmetry-breaking scalar field  
but does not depend on the time and hence on the loop length or Hubble length.
Then, it is vital to note that the loops do not 
exist when the estimated loop length, $l$, is shorter than its thickness, $l<\delta$. 
Moreover, the wavelength of GWs from string loops should be larger than the string thickness,
since the wavelength corresponds to the wavelength of the oscillation on the loop, $\lambda = l/2k$, which also makes sense only when it is longer than $\delta$.
Therefore, we may argue that $k_\mathrm{max}$ is not a constant but loop length dependent as
\begin{equation}
    k_\mathrm{max} = \frac{l(t)}{2 \delta}.  \label{kmaxdelta}
\end{equation}

If we suppose that the system with the strings whose tension is  $G\mu =10^{-11}$ entered the scaling regime to start to form loops at the temperature $T\sim 10^{10}$ GeV, the 
corresponding redshift is $z_{\text{sc}}\sim10^{23}$, 
and the characteristic frequency, $f_\mathrm{sc}$, is evaluated as
$f_\mathrm{sc}\sim 3\times 10^{13}$ Hz.
In this case, the wavelength of GWs at $z_\mathrm{sc} \sim 10^{23}$ 
that corresponds to $f_\mathrm{sc} \sim 3 \times 10^{13}$ Hz today 
is estimated as $\lambda=1/(z_\mathrm{sc} f) \sim 3.8 \times 10^{-13} \mathrm{GeV}^{-1}$, which is a bit longer than the string thickness, $\delta \simeq (\mu/2\pi)^{1/2} \sim 6.6 \times 10^{-14} \mathrm{GeV}^{-1}$. This suggests that in such parameters
the loops whose length $l$ is larger than $\lambda$ exist, and a harmonic mode on that smaller than $k_\mathrm{max} = l/2\delta$ contributes to the GW at that frequency. 
The description of the transition of the GW spectrum from the plateau to the $f^{-1/3}$ decay at $f\simeq f_\mathrm{sc}$ is marginally valid, but the transition to the $f^{-1}$ decay at higher frequencies would be questionable. However, this is the case with a relatively low string tension and an earlier onset of scaling evolution, whose corresponding characteristic frequency is far beyond those for the ongoing or planned GW experiments such as LVK.

We are interested in the case when the transition of the GW spectrum from the plateau to $f^{-1/3}$ decay comes at a frequency smaller than those for LVK mesaurements, $f_\mathrm{sc} < 10^2$ Hz. 
This can be realized, {\it e.g.}, if the string tension is around $G\mu \sim 10^{-5}$ and the scaling regime starts at $z_\mathrm{sc} \sim 10^{14}$. 
In this case, the transition of the GW spectrum from the plateau to the $f^{-1/3}$ decay  occurs at $f\simeq 2 \times 10^{-2} (z_s/10^{14}) (G \mu/10^{-5})^{-1}$ Hz. 
The wavelength of the GW with $f\sim 10^2$ Hz today at the emission $z\sim 10^{14}$ is $1.5 \times 10^8 \mathrm{GeV}^{-1}$, which is much larger than the string thickness $\delta \sim  6.6 \times 10^{-17} \mathrm{GeV}^{-1}$. In other words, the maximum number of the harmonic mode for the loop that emits the GW with $f\sim 10^2$ Hz as the fundamental mode ($k=1$) is $k_\mathrm{max} \sim 10^{24}$, which suggests that we would identify that the $f^{-1/3}$ decay of the GW spectrum continues up to $f\sim 10^{26}$ Hz. 
For $f \sim 10^{28}$ Hz today, we evaluate that the wavelength at $z\simeq 10^{14}$ is comparable to the string thickness. 
Therefore, we conclude that at the GW frequency today of our interest, the spectrum decays with $f^{-1/3}$ even though the numerical integration tells the break to $f^{-1}$ decay due to the artificially small choice of $k_\mathrm{max}. $

Before proceeding, let us give a supporting discussion following that given in Ref.~\cite{Schmitz:2024gds}. As long as the NG approximation holds, this initial time of the onset of the scaling regime, or equivalently, of the GW emission from the string loop should be the time when the so-called friction-dominated era ends 
$t_\mathrm{fric} \sim t_\mathrm{pl}/(G\mu)^2$ with $t_\mathrm{pl}=G^{1/2}$ being the Planck time if it is later than the time when the distance between the long strings becomes smaller than the Hubble length, $t_\mathrm{sc}$, namely, $t_\mathrm{lf} = \mathrm{max} \{ t_\mathrm{fric}, t_\mathrm{sc}\}$. 
However, as discussed in the above, 
the length of a loop as well as the wavelength of the oscillation on the loop
should be longer than the string thickness. 
Thus, we define a ``present frequency dependent'' initial time for the GWs from string loops. 
For a frequency today, $f$, we can determine the time when its corresponding wavelength 
was the comparable to the string thickness due to the cosmic expansion, $a_\mathrm{min}(f) = a_0f \delta = a_0 t_\mathrm{pl} f \sqrt{2 \pi/G\mu}$. 
If it is later than the $t_\mathrm{lf}$ defined in the above, it is the ``initial time'' for the GW with the frequency, $f$, namely,
\begin{equation}
    a_{\text{ini}} (f)=\text{max}\{a_{\text{min}}(f), a(t_{\text{lf}})\}. 
    \label{Kai's t_ini}
\end{equation}
As long as $a_\mathrm{min} (f) < a_\mathrm{lf}$, we conclude that 
the frequency-dependent initial time does not affect the GW spectrum and 
the $f^{-1/3}$ decay holds at the frequency,
\begin{equation}
    f < \frac{\sqrt{G\mu/2\pi}}{t_\mathrm{pl}} z_\mathrm{sc}^{-1} = 2\times 10^{26} \mathrm{Hz} \left(\frac{G\mu}{10^{-5}}\right)^{1/2} \left(\frac{z_\mathrm{sc}}{10^{14}}\right)^{-1}, 
\end{equation}
which is consistent with the estimate in the above.

\subsection{Prospects for observations of GWB}

Finally, let us examine the observational prospects of the GWs from metastable CSs in the delayed scaling scenario. 
While it has been pointed out that metastable CSs are suitable in explaining the PTA measurements. 
According to Ref.~\cite{NANOGrav:2023hvm}, 
the parameters suitable to explain the PTA measurements are identified as
\begin{equation}
    10^{-8} \lesssim G \mu \lesssim 10^{-5}, \quad 7.7 \lesssim \sqrt{\kappa} \lesssim 8.3. 
\end{equation}
However, CSs with larger tension are severely constrained by the constraints from the LVK measurements (Eq.~\eqref{LVKconst}). 
In the delayed scaling scenario, as we have seen, the latter can be avoided, but it is non-trivial if the upper bound of the string tension
is unchanged in this case, since the decay of the GW spectrum at high frequency is relatively mild, $f^{-1/3}$.

\begin{figure}[htbp]
    \centering
    \includegraphics[width=0.9\linewidth]{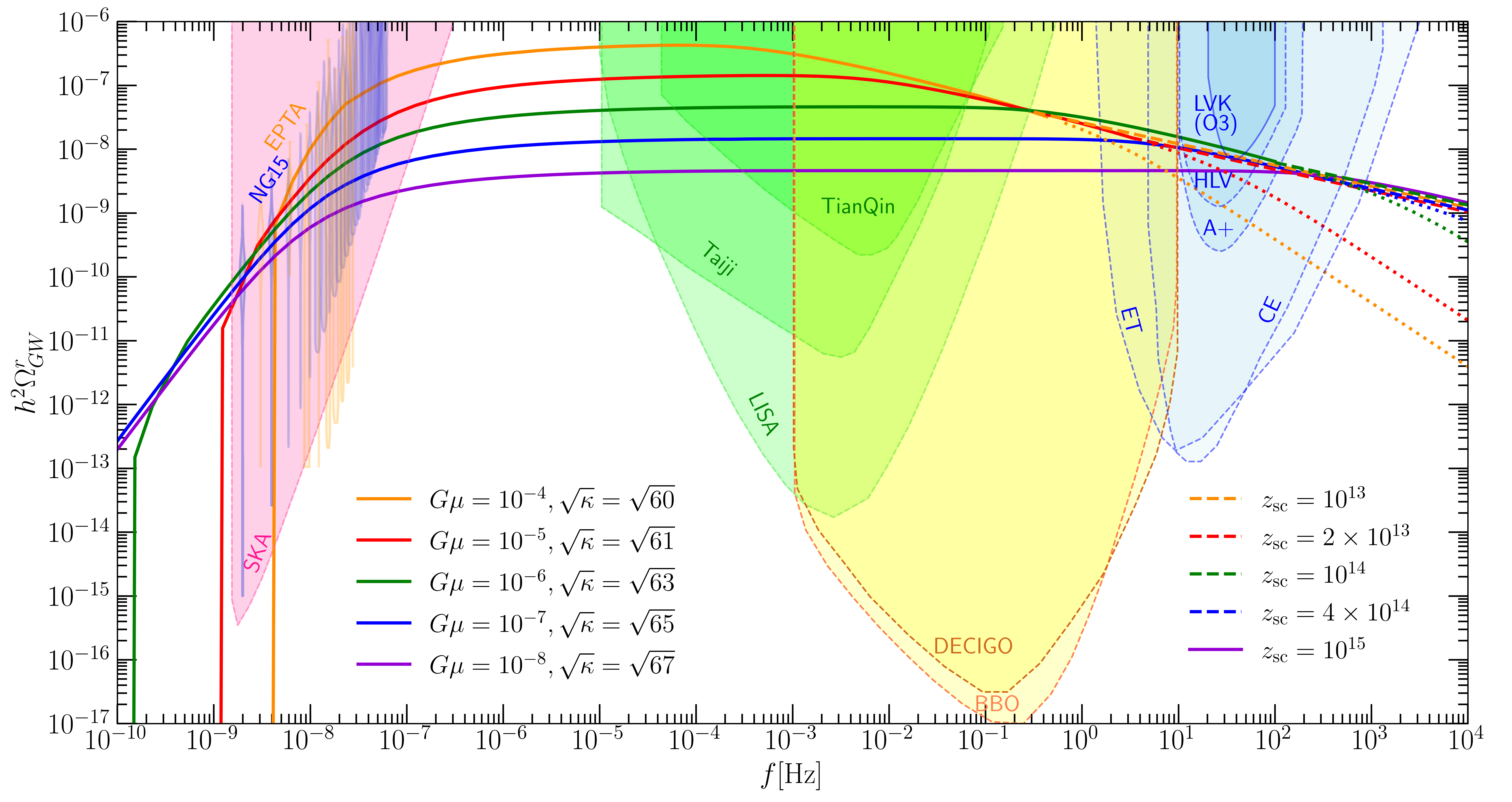}
    \caption{The GW spectra from the metastable CSs in the delayed scaling that are suitable to explain the PTA measurements by merely satisfying the LVK constraints are shown. 
    The string tension, string decay parameter $\kappa$, and the redshift at the onset of the scaling is chosen as $(G\mu, \sqrt{\kappa}, z_\mathrm{sc}) = (10^{-4}, \sqrt{60}, 10^{13})$ [yellow curve], $(10^{-5}, \sqrt{61}, 2 \times 10^{13})$ [red curve], $(10^{-6}, \sqrt{63}, 10^{14})$ [green curve],
    $(10^{-7}, \sqrt{65}, 4 \times 10^{14})$ [blue curve], $(10^{-8}, \sqrt{67}, 10^{15})$ [purple curve]. 
    Solid curves are the numerical results up to $k_\mathrm{max} f_\mathrm{sc}$, while dotted curves are those from $k_\mathrm{max} f_\mathrm{sc}$. 
    Dashed lines are the extrapolation of the $f^{-/13}$ decay at $f>k_\mathrm{max} f_\mathrm{sc}$, which should be physical. 
    Shaded areas indicate the data and sensitivity of the current and future experiments such as NANOGrav 15 year data (NG15)~\cite{NANOGrav:2023gor}, EPTA~\cite{EPTA:2023fyk}, SKA~\cite{Janssen:2014dka}, LISA~\cite{LISA:2017pwj}, Taiji~\cite{Hu:2017mde}, TianQin~\cite{TianQin:2015yph}, DECIGO~\cite{Kawamura:2020pcg}, BBO~\cite{Harry:2006fi},  LVK~\cite{KAGRA:2021kbb} with their updates (HLV, A+), 
    Cosmic Explorer (CE)~\cite{LIGOScientific:2016wof}, and Einstein Telescope (ET)~\cite{Hild:2008ng}. 
    In drawing the sensitivity regions, we used the pubic code available at Ref.~\cite{Fu:2024rsm}. }
    \label{FigObservationalProspects}
\end{figure}

Figure~\ref{FigObservationalProspects} shows the GW spectra with various choices of the parameters, 
$G\mu=10^{-4}, 10^{-5}$, $10^{-6}, 10^{-7}, 10^{-8}$ with 
$\sqrt{\kappa} = \sqrt{60}, \sqrt{61}, \sqrt{63}, \sqrt{65}, \sqrt{67} $ and $z_\mathrm{sc} = 10^{13}, 2 \times 10^{13}, 10^{14}, 4 \times 10^{14}, 10^{15}$, respectively, 
such that at a frequency around $10^{-9} \mathrm{Hz} <f< 10^{-7} \mathrm{Hz}$ the GW spectra reach the PTA measurements while at a higher frequency $f \sim 10$ Hz they merely satisfy the LVK constraints. 
Numerically we have the spectral break at $k_\mathrm{max} f_\mathrm{sc}$ due to the artificially chosen $k_\mathrm{max} = {10^4}$, but as has been discussed in the above, we identify that 
the $f^{-1/3}$ decay continues to much higher frequencies. 
While the spectral shapes are the same at the PTA and LVK frequency ranges in these parameter sets,
the spectral features at the intermediate frequencies are different, which are accessible by future observations at such as LISA, Taiji, and TianQin at mHz frequency ranges or DECIGO or BBO at 1 Hz frequency ranges. The spectral break from the plateau to $f^{1/3}$ decay can be probed by them,  
which can be the smoking gun of this scenario and even determine the parameters of two symmetry breaking scales as well as the dilution during inflation.

We find that the low energy cutoff, $f_\mathrm{lc}$, comes at the frequency larger than that for the PTA measurements, $10^{-8}$ Hz, when $G\mu \gtrsim 10^{-4}$, as has also been pointed out in Ref.~\cite{NANOGrav:2023hvm}. 
This rules out the possibility for such high string tension to explain the PTA measurements. 
Then the upper bound of the string tension is obtained as
\begin{equation}
    G\mu \lesssim 3 \times 10^{-5}, 
\end{equation}
which is unchanged from the constraint given in Ref.~\cite{NANOGrav:2023hvm}. 
This is because it is enough to have the spectral break at $f_\mathrm{low}\simeq 10^{-4}$ Hz, which is much higher than the onset of the plateau, for 
the high frequency tail proportional to $f^{-1/3}$ to come just below the LVK constraint.

\section{Implication to inflation and grand unified theories} \label{sec:implication}

Before concluding, we discuss the implication on inflation and GUTs in the context of the GWB observations and delayed scaling scenario. 
We have seen that $G\mu \simeq 10^{-5}$,  $\sqrt{\kappa} \simeq \sqrt{61} \simeq 7.81$ (or $z_s \sim 10^{10}$), and $z_\mathrm{sc} \sim 2 \times 10^{13}$ are good parameters to explain the PTA measurements also consistent with the constraint by LVK measurements (Eq.~\eqref{LVKconst}). The string tension corresponds to the symmetry breaking scale $v_\mathrm{cs} \sim 4 \times 10^{16} \mathrm{GeV} (G\mu/10^{-5})^{1/2}$. 
The latter corresponds to the monopole mass $m_M\simeq 3 \times 10^{17} \mathrm{GeV} \times (\sqrt{\kappa}/7.8) (G \mu/10^{-5})^{1/2} $ in the semiclassical approximation~\cite{Vilenkin:1982hm,Preskill:1992ck} while recent study suggests a smaller hierarchy~\cite{Chitose:2023dam}. 
Noting that the monopole mass is expressed in terms of the gauge coupling $g_{\cal F}$ and symmetry breaking scale $v_\mathrm{mp}$ as
$m_M \simeq 4 \pi v_\mathrm{mp}/g_{\cal F}$~\cite{Vilenkin:2000jqa}, 
the two symmetry breaking scale should be almost the same, $v_\mathrm{cs} \sim v_\mathrm{mp}$.  
Thus, as has been discussed, we need to have a careful setup such that monopoles are sufficiently diluted before the CS formation. 

Spontaneous symmetry breaking during inflation is often triggered by the Hubble induced mass that appears in models based on supergravity~\cite{Dine:1995uk,Dine:1995kz} or with non-minimal coupling of symmetry breaking Higgs field to Ricci scalar~\cite{Allen:1983dg,Ishikawa:1983kz}. However, the symmetry breaking scale suggested by the PTA measurements, ${\cal O} (10^{16} \mathrm{GeV})$, is much higher than the upper bound of the Hubble parameter during inflation obtained by the measurements of CMB, ${\cal O} (10^{14} \mathrm{GeV})$~\cite{Planck:2018jri}. 
Therefore the Hubble induced mass is not enough to restore the symmetry during inflation unless the symmetry breaking Higgs mass is significantly smaller than the symmetry breaking scale $v_\mathrm{cs}$. Therefore, a direct coupling between inflaton and symmetry breaking Higgs field is a more plausible model to realize the scenario. 

Let us give a concrete model that realizes our scenario as a working example. We here consider the action of inflaton $\phi$, symmetry breaking Higgs field for CS $h$, and that for monopoles $\Phi$ as
\begin{align}
    {\cal L} = &-\frac{1}{2} (\partial_\mu \phi)^2 - |\partial_\mu h|^2 - |\partial_\mu \Phi|^2 \notag \\
    &- V(\phi) - \lambda_h (|h|^2 - v_\mathrm{cs})^2 - \lambda_\Phi   (|\Phi|^2 - v_\mathrm{mp})^2-\lambda_{\phi h} \phi^2 |h|^2 - \lambda_{\phi \Phi} \phi^2 |\Phi|^2,
\end{align}
where we consider the $R^2$-inflation like potential~\cite{Starobinsky:1980te,Barrow:1988xh},
\begin{equation}
    V(\phi) = \frac{3}{4} M^2 M_\mathrm{pl}^2 \left(1 - \exp\left[-\sqrt{\frac{2}{3}} \frac{\phi}{M_\mathrm{pl}} \right] \right)^2,
\end{equation}
where $M_\mathrm{pl} =(8\pi G)^{-1/2} \simeq 2.43 \times 10^{18}$ GeV is the reduced Planck mass 
and $M\simeq 3 \times 10^{13}$ GeV is the scalaron mass with which the density perturbation generated in this model fits the present cosmological observations~\cite{Planck:2018jri}. 
We require $\lambda_h, \lambda_\Phi  \lesssim 10^{-4}$ such that the potential energy of the symmetry breaking scalars do not spoil inflation. 
In this setup, the inflaton field $\phi$ start rolling down the potential $V(\phi)$ from a large field value $\phi \gg M_\mathrm{pl}$. 
The monopole forming symmetry breaking takes place when the induced mass becomes comparable to the tachyonic mass, $\phi \simeq  \sqrt{\lambda_{\Phi}/\lambda_{\phi \Phi}}v_\mathrm{mp} $, 
while the string forming one does at $\phi \simeq \sqrt{\lambda_h/\lambda_{\phi h}}v_\mathrm{cs}$. 
Even though $v_\mathrm{cs} \sim v_\mathrm{mp}$, if we have a hierarchy $\lambda_\Phi/\lambda_{\phi\Phi} \gg \lambda_h/\lambda_{\phi h}$, we can have a sufficient period of the dilution of monopoles. 
Inflation ends when the slow-roll parameter $\epsilon \equiv (M_\mathrm{pl}^2/2) (\partial_\phi V /V)^2$ becomes order of unity, $\phi = \sqrt{3/2} \ln (1+2/\sqrt{3}) M_\mathrm{pl} \simeq 0.94 M_\mathrm{pl} $. 
Note that if we have another hierarchy between the couplings, $\lambda_\Phi \gg \lambda_{\phi \Phi}, \lambda_h \gg\lambda_{\phi h}$, symmetry breakings can take place during inflation.

Now we examine the time when the CS network enters the scaling regime in this setup. 
Taking the separation of the long string just after the symmetry breaking as $L_\mathrm{sb} = \zeta H_\mathrm{inf}^{-1} \simeq 2 \zeta M^{-1}$ and the number of $e$-folds between the symmetry breaking and inflation end as $\Delta N_\mathrm{cs}$, the separation at the end of inflation is estimated as $L_\mathrm{ie} = 2 e^{\Delta N_\mathrm{cs}} \zeta M^{-1}$. 
Supposing the instant reheating, during radiation dominated era it is evaluated as
\begin{equation}
    L(z) = 2 e^{\Delta N_\mathrm{cs}} \zeta M^{-1} \frac{a(z)}{a_\mathrm{ie}} = 2 e^{\Delta N_\mathrm{cs}} \zeta M^{-1} \frac{T_\mathrm{re}}{T(z)}, 
\end{equation}
where $a_\mathrm{ie}$ the scale factor at the end of inflation and $T_\mathrm{re} = (45/2 \pi^2 g_*)^{1/4} \sqrt{M M_\mathrm{pl}}$ is the reheating temperature with $g_*^\mathrm{re}=106.75$ being the number of relativistic particles at reheating. 
For concreteness, we take an ansatz that the system enters the scaling regime and starts to emit the GWs when the long loop separation becomes that in the scaling regime for the VOS model, $L(t) = \xi_r t = \xi_r/2H$ with $\xi_r= 0.27$~\cite{Kibble:1984hp,Martins:1996jp,Martins:2000cs}. 
The redshift when the system starts to emit GWs is then given by
\begin{align}
    z_\mathrm{sc} &\simeq \left(\frac{45}{2 \pi^2 g_*}\right)^{1/4} \left(\frac{g_{*s}(z_\mathrm{sc})}{g_{*s}^0}\right)^{1/3}  e^{-\Delta N_\mathrm{cs}} \frac{\xi_r}{\zeta} \frac{\sqrt{M M_\mathrm{pl}}}{T_0} \notag \\
    &\simeq 2 \times 10^{13} \zeta^{-1} \left( \frac{e^{-\Delta N_\mathrm{cs}}}{10^{-15}}\right) \left(\frac{\xi_r}{0.27} \right)\left(\frac{M}{3\times 10^{13} \mathrm{GeV}} \right)^{1/2},  
\end{align}
where $g_{*s}$ is the effective number of relativistic degree of freedom for entropy where we have taken $g_{*s}^0 = 3.91$ today and $g_{*s}^{\mathrm{sc}}=g_{*}^{\mathrm{sc}}=106.75$, and $T_0=2.4 \times 10^{-13}$ GeV is the present temperature. 
Noting that the number of $e$-folds from the time when the inflaton field value is $\phi$ to inflation end is given as $N_\mathrm{e}\simeq (3/4) \exp [\sqrt{2/3} \phi/M_\mathrm{pl}]$, we conclude that $z_\mathrm{sc} \simeq 10^{13}$ is realized for $\phi_\mathrm{cs} \simeq 4.7 M_\mathrm{pl}$ and hence $\lambda_h/\lambda_{\phi h} \sim 10^4 (v_\mathrm{cs}/10^{16} \mathrm{GeV})^{-2}$. The monopoles are sufficiently diluted if $\lambda_\Phi/\lambda_{\phi \Phi}$ is much larger. 
Smaller coupling between the symmetry breaking fields and inflaton, $\lambda_{\phi h}$ and $\lambda_{\phi \Phi}$, are favored, which is preferable to suppresses the radiative corrections to the inflaton potential.

\section{Summary} \label{sec:summary}

In this study, we studied the GW spectrum from the metastable CSs in the delayed scaling scenario. 
The metastable CSs are plausible GW source that explain the PTA measurements recently reported, if the CSs decay through the monopole-antimonopole pair creation around the redshift $z_s \simeq 10^{10}$.  However, they are subject to severe constraints from the non-detection of stochastic GWB at LVK measurements. 
While the constraints from LVK measurements are unavoidable for CSs that are formed through the usual Kibble-Zurek mechanism triggered by thermal potential, they can be avoided if we consider the delayed scaling scenario, where the CS network enters the scaling regime to emit GWB much later than the time of the long string formation. 
If the symmetry breaking takes place during inflation, this scenario is naturally realized. 
Note that in this case, monopoles, whose energy scale should be relatively close to those for CSs, can be sufficiently diluted such that the CS network can once form. 

We identified the feature of the GW spectrum in this scenario; in particular, at high frequency we identify a $f^{-1/3}$ decay, dominated by higher harmonic modes, which is the same to the case with the stable CS network in the delayed scaling scenario. 
Note that GWB from CS network at high frequencies has contributions from all the string loop in the cosmic history, and hence it is non-trivial if the frequency of the spectral break and the slope is unchanged compared to the stable CS network case. 
In the literature, transition to $f^{-1}$ decay is discussed due to the upper limit of the harmonic mode of the loop oscillation. 
Supposing that the upper limit comes from the string thickness with which the NG approximation breaks down, 
we showed that such transition does not appear at the frequency range and the detectable amplitude at the existing or planned GW observations. 
Our scenario can be thus probed by the detection of the $f^{-1/3}$ decay of the stochastic GWB at frequency around $10^2$ Hz (for LVK) or lower (for such as LISA, Taiji, and TianQin or DECIGO and BBO), together with the spectral break from the plateau to the $f^{-1/3}$ decay . 
The CS tension can be up to $G\mu \simeq 3 \times 10^{-5}$ to explain the PTA measurements if the CS decay rate is around $\sqrt{\kappa} \simeq \sqrt{61}$ or $z_s \simeq 10^{10}$, which is consistent with the report by NANOGrav~\cite{NANOGrav:2023hvm},  while merely avoiding the LVK constraints~\cite{KAGRA:2021kbb} if the CS network enters the scaling regime around $z_\mathrm{sc}\simeq 2 \times 10^{13}$ or later. 
This is comparable to the upper bound reported in NANOGrav 15 year data~\cite{NANOGrav:2023gor} without taking into account the high-frequency constraints. 
Since the decay at high frequency is milder, $f^{-1/3}$, 
it was non-trivial if this constraints holds in the delayed scaling scenario. 
We identified that it holds because of a sufficiently large redshift of the onset of the scaling regime required to satisfy the LVK constraint. 

We also exhibited a toy inflationary model that realizes the scenario. 
In this model, the symmetry breaking is triggered by the direct coupling between the inflaton and symmetry breaking fields. 
By tuning the scalar four point couplings, we show that desired cosmic history, delayed scaling of CS network at $z_\mathrm{sc} \sim 2 \times 10^{13}$ with sufficiently dilution of monopoles, is realized. 
However, it is not clear if such a small couplings can be protected against the quantum corrections. 
More realistic model building including more precise evaluation how the system enters the scaling regime and observational prospects is left for future study.

\acknowledgments
Y.~H. would like to thank Ye-ling Zhou for useful discussions and comments. 
K.~K. would like to thank Nagoya University for kind hospitality during the completion of the present work. 
The work of K.~K. was supported by the National Natural Science Foundation of China (NSFC) under Grant No.~12347103 and JSPS KAKENHI Grant-in-Aid for Challenging Research (Exploratory) JP23K17687.

\bibliographystyle{utphys}
\bibliography{ref}

\providecommand{\href}[2]{#2}\begingroup\raggedright\begin{thebibliography}{100}

\bibitem{LIGOScientific:2014pky}
{\bfseries LIGO Scientific} Collaboration, J.~Aasi {\em et~al.}, ``{Advanced
  LIGO},'' \href{http://dx.doi.org/10.1088/0264-9381/32/7/074001}{{\em Class.
  Quant. Grav.} {\bfseries 32} (2015) 074001},
  \href{http://arxiv.org/abs/1411.4547}{{\ttfamily arXiv:1411.4547 [gr-qc]}}.

\bibitem{VIRGO:2014yos}
{\bfseries VIRGO} Collaboration, F.~Acernese {\em et~al.}, ``{Advanced Virgo: a
  second-generation interferometric gravitational wave detector},''
  \href{http://dx.doi.org/10.1088/0264-9381/32/2/024001}{{\em Class. Quant.
  Grav.} {\bfseries 32} no.~2, (2015) 024001},
  \href{http://arxiv.org/abs/1408.3978}{{\ttfamily arXiv:1408.3978 [gr-qc]}}.

\bibitem{Somiya:2011np}
{\bfseries KAGRA} Collaboration, K.~Somiya, ``{Detector configuration of KAGRA:
  The Japanese cryogenic gravitational-wave detector},''
  \href{http://dx.doi.org/10.1088/0264-9381/29/12/124007}{{\em Class. Quant.
  Grav.} {\bfseries 29} (2012) 124007},
  \href{http://arxiv.org/abs/1111.7185}{{\ttfamily arXiv:1111.7185 [gr-qc]}}.

\bibitem{Aso:2013eba}
{\bfseries KAGRA} Collaboration, Y.~Aso, Y.~Michimura, K.~Somiya, M.~Ando,
  O.~Miyakawa, T.~Sekiguchi, D.~Tatsumi, and H.~Yamamoto, ``{Interferometer
  design of the KAGRA gravitational wave detector},''
  \href{http://dx.doi.org/10.1103/PhysRevD.88.043007}{{\em Phys. Rev. D}
  {\bfseries 88} no.~4, (2013) 043007},
  \href{http://arxiv.org/abs/1306.6747}{{\ttfamily arXiv:1306.6747 [gr-qc]}}.

\bibitem{LISA:2017pwj}
{\bfseries LISA} Collaboration, P.~Amaro-Seoane {\em et~al.}, ``{Laser
  Interferometer Space Antenna},''
  \href{http://arxiv.org/abs/1702.00786}{{\ttfamily arXiv:1702.00786
  [astro-ph.IM]}}.

\bibitem{Hu:2017mde}
W.-R. Hu and Y.-L. Wu, ``{The Taiji Program in Space for gravitational wave
  physics and the nature of gravity},''
  \href{http://dx.doi.org/10.1093/nsr/nwx116}{{\em Natl. Sci. Rev.} {\bfseries
  4} no.~5, (2017) 685--686}.

\bibitem{Ruan:2018tsw}
W.-H. Ruan, Z.-K. Guo, R.-G. Cai, and Y.-Z. Zhang, ``{Taiji program:
  Gravitational-wave sources},''
  \href{http://dx.doi.org/10.1142/S0217751X2050075X}{{\em Int. J. Mod. Phys. A}
  {\bfseries 35} no.~17, (2020) 2050075},
  \href{http://arxiv.org/abs/1807.09495}{{\ttfamily arXiv:1807.09495 [gr-qc]}}.

\bibitem{TianQin:2015yph}
{\bfseries TianQin} Collaboration, J.~Luo {\em et~al.}, ``{TianQin: a
  space-borne gravitational wave detector},''
  \href{http://dx.doi.org/10.1088/0264-9381/33/3/035010}{{\em Class. Quant.
  Grav.} {\bfseries 33} no.~3, (2016) 035010},
  \href{http://arxiv.org/abs/1512.02076}{{\ttfamily arXiv:1512.02076
  [astro-ph.IM]}}.

\bibitem{TianQin:2020hid}
{\bfseries TianQin} Collaboration, J.~Mei {\em et~al.}, ``{The TianQin project:
  current progress on science and technology},''
  \href{http://dx.doi.org/10.1093/ptep/ptaa114}{{\em PTEP} {\bfseries 2021}
  no.~5, (2021) 05A107}, \href{http://arxiv.org/abs/2008.10332}{{\ttfamily
  arXiv:2008.10332 [gr-qc]}}.

\bibitem{Seto:2001qf}
N.~Seto, S.~Kawamura, and T.~Nakamura, ``{Possibility of direct measurement of
  the acceleration of the universe using 0.1-Hz band laser interferometer
  gravitational wave antenna in space},''
  \href{http://dx.doi.org/10.1103/PhysRevLett.87.221103}{{\em Phys. Rev. Lett.}
  {\bfseries 87} (2001) 221103},
  \href{http://arxiv.org/abs/astro-ph/0108011}{{\ttfamily
  arXiv:astro-ph/0108011}}.

\bibitem{Corbin:2005ny}
V.~Corbin and N.~J. Cornish, ``{Detecting the cosmic gravitational wave
  background with the big bang observer},''
  \href{http://dx.doi.org/10.1088/0264-9381/23/7/014}{{\em Class. Quant. Grav.}
  {\bfseries 23} (2006) 2435--2446},
  \href{http://arxiv.org/abs/gr-qc/0512039}{{\ttfamily arXiv:gr-qc/0512039}}.

\bibitem{Planck:2018jri}
{\bfseries Planck} Collaboration, Y.~Akrami {\em et~al.}, ``{Planck 2018
  results. X. Constraints on inflation},''
  \href{http://dx.doi.org/10.1051/0004-6361/201833887}{{\em Astron. Astrophys.}
  {\bfseries 641} (2020) A10},
  \href{http://arxiv.org/abs/1807.06211}{{\ttfamily arXiv:1807.06211
  [astro-ph.CO]}}.

\bibitem{LiteBIRD:2022cnt}
{\bfseries LiteBIRD} Collaboration, E.~Allys {\em et~al.}, ``{Probing Cosmic
  Inflation with the LiteBIRD Cosmic Microwave Background Polarization
  Survey},'' \href{http://dx.doi.org/10.1093/ptep/ptac150}{{\em PTEP}
  {\bfseries 2023} no.~4, (2023) 042F01},
  \href{http://arxiv.org/abs/2202.02773}{{\ttfamily arXiv:2202.02773
  [astro-ph.IM]}}.

\bibitem{Abazajian:2019eic}
K.~Abazajian {\em et~al.}, ``{CMB-S4 Science Case, Reference Design, and
  Project Plan},'' \href{http://arxiv.org/abs/1907.04473}{{\ttfamily
  arXiv:1907.04473 [astro-ph.IM]}}.

\bibitem{NANOGrav:2023gor}
{\bfseries NANOGrav} Collaboration, G.~Agazie {\em et~al.}, ``{The NANOGrav 15
  yr Data Set: Evidence for a Gravitational-wave Background},''
  \href{http://dx.doi.org/10.3847/2041-8213/acdac6}{{\em Astrophys. J. Lett.}
  {\bfseries 951} no.~1, (2023) L8},
  \href{http://arxiv.org/abs/2306.16213}{{\ttfamily arXiv:2306.16213
  [astro-ph.HE]}}.

\bibitem{EPTA:2023fyk}
{\bfseries EPTA, InPTA:} Collaboration, J.~Antoniadis {\em et~al.}, ``{The
  second data release from the European Pulsar Timing Array - III. Search for
  gravitational wave signals},''
  \href{http://dx.doi.org/10.1051/0004-6361/202346844}{{\em Astron. Astrophys.}
  {\bfseries 678} (2023) A50},
  \href{http://arxiv.org/abs/2306.16214}{{\ttfamily arXiv:2306.16214
  [astro-ph.HE]}}.

\bibitem{Reardon:2023gzh}
D.~J. Reardon {\em et~al.}, ``{Search for an Isotropic Gravitational-wave
  Background with the Parkes Pulsar Timing Array},''
  \href{http://dx.doi.org/10.3847/2041-8213/acdd02}{{\em Astrophys. J. Lett.}
  {\bfseries 951} no.~1, (2023) L6},
  \href{http://arxiv.org/abs/2306.16215}{{\ttfamily arXiv:2306.16215
  [astro-ph.HE]}}.

\bibitem{Xu:2023wog}
H.~Xu {\em et~al.}, ``{Searching for the Nano-Hertz Stochastic Gravitational
  Wave Background with the Chinese Pulsar Timing Array Data Release I},''
  \href{http://dx.doi.org/10.1088/1674-4527/acdfa5}{{\em Res. Astron.
  Astrophys.} {\bfseries 23} no.~7, (2023) 075024},
  \href{http://arxiv.org/abs/2306.16216}{{\ttfamily arXiv:2306.16216
  [astro-ph.HE]}}.

\bibitem{NANOGrav:2023hvm}
{\bfseries NANOGrav} Collaboration, A.~Afzal {\em et~al.}, ``{The NANOGrav 15
  yr Data Set: Search for Signals from New Physics},''
  \href{http://dx.doi.org/10.3847/2041-8213/acdc91}{{\em Astrophys. J. Lett.}
  {\bfseries 951} no.~1, (2023) L11},
  \href{http://arxiv.org/abs/2306.16219}{{\ttfamily arXiv:2306.16219
  [astro-ph.HE]}}. [Erratum: Astrophys.J.Lett. 971, L27 (2024), Erratum:
  Astrophys.J. 971, L27 (2024)].

\bibitem{Vilenkin:1981bx}
A.~Vilenkin, ``{Gravitational radiation from cosmic strings},''
  \href{http://dx.doi.org/10.1016/0370-2693(81)91144-8}{{\em Phys. Lett. B}
  {\bfseries 107} (1981) 47--50}.

\bibitem{Hogan:1984is}
C.~J. Hogan and M.~J. Rees, ``{Gravitational interactions of cosmic strings},''
  \href{http://dx.doi.org/10.1038/311109a0}{{\em Nature} {\bfseries 311} (1984)
  109--113}.

\bibitem{Vachaspati:1984gt}
T.~Vachaspati and A.~Vilenkin, ``{Gravitational Radiation from Cosmic
  Strings},'' \href{http://dx.doi.org/10.1103/PhysRevD.31.3052}{{\em Phys. Rev.
  D} {\bfseries 31} (1985) 3052}.

\bibitem{Accetta:1988bg}
F.~S. Accetta and L.~M. Krauss, ``{The stochastic gravitational wave spectrum
  resulting from cosmic string evolution},''
  \href{http://dx.doi.org/10.1016/0550-3213(89)90628-7}{{\em Nucl. Phys. B}
  {\bfseries 319} (1989) 747--764}.

\bibitem{Vilenkin:1982hm}
A.~Vilenkin, ``{COSMOLOGICAL EVOLUTION OF MONOPOLES CONNECTED BY STRINGS},''
  \href{http://dx.doi.org/10.1016/0550-3213(82)90037-2}{{\em Nucl. Phys. B}
  {\bfseries 196} (1982) 240--258}.

\bibitem{Hindmarsh:2011qj}
M.~Hindmarsh, ``{Signals of Inflationary Models with Cosmic Strings},''
  \href{http://dx.doi.org/10.1143/PTPS.190.197}{{\em Prog. Theor. Phys. Suppl.}
  {\bfseries 190} (2011) 197--228},
  \href{http://arxiv.org/abs/1106.0391}{{\ttfamily arXiv:1106.0391
  [astro-ph.CO]}}.

\bibitem{Buchmuller:2013lra}
W.~Buchm\"uller, V.~Domcke, K.~Kamada, and K.~Schmitz, ``{The Gravitational
  Wave Spectrum from Cosmological $B-L$ Breaking},''
  \href{http://dx.doi.org/10.1088/1475-7516/2013/10/003}{{\em JCAP} {\bfseries
  10} (2013) 003}, \href{http://arxiv.org/abs/1305.3392}{{\ttfamily
  arXiv:1305.3392 [hep-ph]}}.

\bibitem{Auclair:2019wcv}
P.~Auclair {\em et~al.}, ``{Probing the gravitational wave background from
  cosmic strings with LISA},''
  \href{http://dx.doi.org/10.1088/1475-7516/2020/04/034}{{\em JCAP} {\bfseries
  04} (2020) 034}, \href{http://arxiv.org/abs/1909.00819}{{\ttfamily
  arXiv:1909.00819 [astro-ph.CO]}}.

\bibitem{Gouttenoire:2019kij}
Y.~Gouttenoire, G.~Servant, and P.~Simakachorn, ``{Beyond the Standard Models
  with Cosmic Strings},''
  \href{http://dx.doi.org/10.1088/1475-7516/2020/07/032}{{\em JCAP} {\bfseries
  07} (2020) 032}, \href{http://arxiv.org/abs/1912.02569}{{\ttfamily
  arXiv:1912.02569 [hep-ph]}}.

\bibitem{Binetruy:2012ze}
P.~Binetruy, A.~Bohe, C.~Caprini, and J.-F. Dufaux, ``{Cosmological Backgrounds
  of Gravitational Waves and eLISA/NGO: Phase Transitions, Cosmic Strings and
  Other Sources},'' \href{http://dx.doi.org/10.1088/1475-7516/2012/06/027}{{\em
  JCAP} {\bfseries 06} (2012) 027},
  \href{http://arxiv.org/abs/1201.0983}{{\ttfamily arXiv:1201.0983 [gr-qc]}}.

\bibitem{Sousa:2020sxs}
L.~Sousa, P.~P. Avelino, and G.~S.~F. Guedes, ``{Full analytical approximation
  to the stochastic gravitational wave background generated by cosmic string
  networks},'' \href{http://dx.doi.org/10.1103/PhysRevD.101.103508}{{\em Phys.
  Rev. D} {\bfseries 101} no.~10, (2020) 103508},
  \href{http://arxiv.org/abs/2002.01079}{{\ttfamily arXiv:2002.01079
  [astro-ph.CO]}}.

\bibitem{Blanco-Pillado:2024aca}
{\bfseries LISA Cosmology Working Group} Collaboration, J.~J. Blanco-Pillado,
  Y.~Cui, S.~Kuroyanagi, M.~Lewicki, G.~Nardini, M.~Pieroni, I.~Y. Rybak,
  L.~Sousa, and J.~M. Wachter, ``{Gravitational waves from cosmic strings in
  LISA: reconstruction pipeline and physics interpretation},''
  \href{http://arxiv.org/abs/2405.03740}{{\ttfamily arXiv:2405.03740
  [astro-ph.CO]}}.

\bibitem{Schmitz:2024gds}
K.~Schmitz and T.~Schr\"oder, ``{Gravitational waves from cosmic strings for
  pedestrians},'' \href{http://arxiv.org/abs/2412.20907}{{\ttfamily
  arXiv:2412.20907 [astro-ph.CO]}}.

\bibitem{Nielsen:1973cs}
H.~B. Nielsen and P.~Olesen, ``{Vortex Line Models for Dual Strings},''
  \href{http://dx.doi.org/10.1016/0550-3213(73)90350-7}{{\em Nucl. Phys. B}
  {\bfseries 61} (1973) 45--61}.

\bibitem{Hindmarsh:1994re}
M.~B. Hindmarsh and T.~W.~B. Kibble, ``{Cosmic strings},''
  \href{http://dx.doi.org/10.1088/0034-4885/58/5/001}{{\em Rept. Prog. Phys.}
  {\bfseries 58} (1995) 477--562},
  \href{http://arxiv.org/abs/hep-ph/9411342}{{\ttfamily arXiv:hep-ph/9411342}}.

\bibitem{Vilenkin:2000jqa}
A.~Vilenkin and E.~P.~S. Shellard, {\em {Cosmic Strings and Other Topological
  Defects}}.
\newblock Cambridge University Press, 7, 2000.

\bibitem{Kibble:1976sj}
T.~W.~B. Kibble, ``{Topology of Cosmic Domains and Strings},''
  \href{http://dx.doi.org/10.1088/0305-4470/9/8/029}{{\em J. Phys. A}
  {\bfseries 9} (1976) 1387--1398}.

\bibitem{Kibble:1980mv}
T.~W.~B. Kibble, ``{Some Implications of a Cosmological Phase Transition},''
  \href{http://dx.doi.org/10.1016/0370-1573(80)90091-5}{{\em Phys. Rept.}
  {\bfseries 67} (1980) 183}.

\bibitem{Zurek:1985qw}
W.~H. Zurek, ``{Cosmological Experiments in Superfluid Helium?},''
  \href{http://dx.doi.org/10.1038/317505a0}{{\em Nature} {\bfseries 317} (1985)
  505--508}.

\bibitem{Kibble:1984hp}
T.~W.~B. Kibble, ``{Evolution of a system of cosmic strings},''
  \href{http://dx.doi.org/10.1016/0550-3213(85)90596-6}{{\em Nucl. Phys. B}
  {\bfseries 252} (1985) 227}. [Erratum: Nucl.Phys.B 261, 750 (1985)].

\bibitem{Albrecht:1989mk}
A.~Albrecht and N.~Turok, ``{Evolution of Cosmic String Networks},''
  \href{http://dx.doi.org/10.1103/PhysRevD.40.973}{{\em Phys. Rev. D}
  {\bfseries 40} (1989) 973--1001}.

\bibitem{Jeannerot:2003qv}
R.~Jeannerot, J.~Rocher, and M.~Sakellariadou, ``{How generic is cosmic string
  formation in SUSY GUTs},''
  \href{http://dx.doi.org/10.1103/PhysRevD.68.103514}{{\em Phys. Rev. D}
  {\bfseries 68} (2003) 103514},
  \href{http://arxiv.org/abs/hep-ph/0308134}{{\ttfamily arXiv:hep-ph/0308134}}.

\bibitem{Leblond:2009fq}
L.~Leblond, B.~Shlaer, and X.~Siemens, ``{Gravitational Waves from Broken
  Cosmic Strings: The Bursts and the Beads},''
  \href{http://dx.doi.org/10.1103/PhysRevD.79.123519}{{\em Phys. Rev. D}
  {\bfseries 79} (2009) 123519},
  \href{http://arxiv.org/abs/0903.4686}{{\ttfamily arXiv:0903.4686
  [astro-ph.CO]}}.

\bibitem{Buchmuller:2019gfy}
W.~Buchmuller, V.~Domcke, H.~Murayama, and K.~Schmitz, ``{Probing the scale of
  grand unification with gravitational waves},''
  \href{http://dx.doi.org/10.1016/j.physletb.2020.135764}{{\em Phys. Lett. B}
  {\bfseries 809} (2020) 135764},
  \href{http://arxiv.org/abs/1912.03695}{{\ttfamily arXiv:1912.03695
  [hep-ph]}}.

\bibitem{Buchmuller:2020lbh}
W.~Buchmuller, V.~Domcke, and K.~Schmitz, ``{From NANOGrav to LIGO with
  metastable cosmic strings},''
  \href{http://dx.doi.org/10.1016/j.physletb.2020.135914}{{\em Phys. Lett. B}
  {\bfseries 811} (2020) 135914},
  \href{http://arxiv.org/abs/2009.10649}{{\ttfamily arXiv:2009.10649
  [astro-ph.CO]}}.

\bibitem{Buchmuller:2021mbb}
W.~Buchmuller, V.~Domcke, and K.~Schmitz, ``{Stochastic gravitational-wave
  background from metastable cosmic strings},''
  \href{http://dx.doi.org/10.1088/1475-7516/2021/12/006}{{\em JCAP} {\bfseries
  12} no.~12, (2021) 006}, \href{http://arxiv.org/abs/2107.04578}{{\ttfamily
  arXiv:2107.04578 [hep-ph]}}.

\bibitem{Buchmuller:2023aus}
W.~Buchmuller, V.~Domcke, and K.~Schmitz, ``{Metastable cosmic strings},''
  \href{http://dx.doi.org/10.1088/1475-7516/2023/11/020}{{\em JCAP} {\bfseries
  11} (2023) 020}, \href{http://arxiv.org/abs/2307.04691}{{\ttfamily
  arXiv:2307.04691 [hep-ph]}}.

\bibitem{Chitose:2024pmz}
A.~Chitose, M.~Ibe, S.~Neda, and S.~Shirai, ``{Gravitational Waves from
  Metastable Cosmic Strings in Supersymmetric New Inflation Model},''
  \href{http://arxiv.org/abs/2411.13299}{{\ttfamily arXiv:2411.13299
  [hep-ph]}}.

\bibitem{Preskill:1992ck}
J.~Preskill and A.~Vilenkin, ``{Decay of metastable topological defects},''
  \href{http://dx.doi.org/10.1103/PhysRevD.47.2324}{{\em Phys. Rev. D}
  {\bfseries 47} (1993) 2324--2342},
  \href{http://arxiv.org/abs/hep-ph/9209210}{{\ttfamily arXiv:hep-ph/9209210}}.

\bibitem{Chitose:2023dam}
A.~Chitose, M.~Ibe, Y.~Nakayama, S.~Shirai, and K.~Watanabe, ``{Revisiting
  metastable cosmic string breaking},''
  \href{http://dx.doi.org/10.1007/JHEP04(2024)068}{{\em JHEP} {\bfseries 04}
  (2024) 068}, \href{http://arxiv.org/abs/2312.15662}{{\ttfamily
  arXiv:2312.15662 [hep-ph]}}.

\bibitem{KAGRA:2021kbb}
{\bfseries KAGRA, Virgo, LIGO Scientific} Collaboration, R.~Abbott {\em
  et~al.}, ``{Upper limits on the isotropic gravitational-wave background from
  Advanced LIGO and Advanced Virgo\textquoteright{}s third observing run},''
  \href{http://dx.doi.org/10.1103/PhysRevD.104.022004}{{\em Phys. Rev. D}
  {\bfseries 104} no.~2, (2021) 022004},
  \href{http://arxiv.org/abs/2101.12130}{{\ttfamily arXiv:2101.12130 [gr-qc]}}.

\bibitem{Lazarides:1984pq}
G.~Lazarides and Q.~Shafi, ``{Extended Structures at Intermediate Scales in an
  Inflationary Cosmology},''
  \href{http://dx.doi.org/10.1016/0370-2693(84)91605-8}{{\em Phys. Lett. B}
  {\bfseries 148} (1984) 35--38}.

\bibitem{Shafi:1984tt}
Q.~Shafi and A.~Vilenkin, ``{Spontaneously Broken Global Symmetries and
  Cosmology},'' \href{http://dx.doi.org/10.1103/PhysRevD.29.1870}{{\em Phys.
  Rev. D} {\bfseries 29} (1984) 1870}.

\bibitem{Vishniac:1986sk}
E.~T. Vishniac, K.~A. Olive, and D.~Seckel, ``{Cosmic Strings and Inflation},''
  \href{http://dx.doi.org/10.1016/0550-3213(87)90403-2}{{\em Nucl. Phys. B}
  {\bfseries 289} (1987) 717--734}.

\bibitem{Kofman:1986wm}
L.~A. Kofman and A.~D. Linde, ``{Generation of Density Perturbations in the
  Inflationary Cosmology},''
  \href{http://dx.doi.org/10.1016/0550-3213(87)90698-5}{{\em Nucl. Phys. B}
  {\bfseries 282} (1987) 555}.

\bibitem{Yokoyama:1988zza}
J.~Yokoyama, ``{Natural Way Out of the Conflict Between Cosmic Strings and
  Inflation},'' \href{http://dx.doi.org/10.1016/0370-2693(88)91316-0}{{\em
  Phys. Lett. B} {\bfseries 212} (1988) 273--276}.

\bibitem{Yokoyama:1989pa}
J.~Yokoyama, ``{INFLATION CAN SAVE COSMIC STRINGS},''
  \href{http://dx.doi.org/10.1103/PhysRevLett.63.712}{{\em Phys. Rev. Lett.}
  {\bfseries 63} (1989) 712}.

\bibitem{Nagasawa:1991zr}
M.~Nagasawa and J.~Yokoyama, ``{Phase transitions triggered by quantum
  fluctuations in the inflationary universe},''
  \href{http://dx.doi.org/10.1016/0550-3213(92)90294-L}{{\em Nucl. Phys. B}
  {\bfseries 370} (1992) 472--490}.

\bibitem{Basu:1993rf}
R.~Basu and A.~Vilenkin, ``{Evolution of topological defects during
  inflation},'' \href{http://dx.doi.org/10.1103/PhysRevD.50.7150}{{\em Phys.
  Rev. D} {\bfseries 50} (1994) 7150--7153},
  \href{http://arxiv.org/abs/gr-qc/9402040}{{\ttfamily arXiv:gr-qc/9402040}}.

\bibitem{Freese:1995vp}
K.~Freese, T.~Gherghetta, and H.~Umeda, ``{Moduli inflation with large scale
  structure produced by topological defects},''
  \href{http://dx.doi.org/10.1103/PhysRevD.54.6083}{{\em Phys. Rev. D}
  {\bfseries 54} (1996) 6083--6087},
  \href{http://arxiv.org/abs/hep-ph/9512211}{{\ttfamily arXiv:hep-ph/9512211}}.

\bibitem{Kamada:2012ag}
K.~Kamada, Y.~Miyamoto, and J.~Yokoyama, ``{Evading the pulsar constraints on
  the cosmic string tension in supergravity inflation},''
  \href{http://dx.doi.org/10.1088/1475-7516/2012/10/023}{{\em JCAP} {\bfseries
  10} (2012) 023}, \href{http://arxiv.org/abs/1204.3237}{{\ttfamily
  arXiv:1204.3237 [astro-ph.CO]}}.

\bibitem{Linde:2013aya}
A.~Linde, ``{Chaotic inflation in supergravity and cosmic string production},''
  \href{http://dx.doi.org/10.1103/PhysRevD.88.123503}{{\em Phys. Rev. D}
  {\bfseries 88} (2013) 123503},
  \href{http://arxiv.org/abs/1303.4435}{{\ttfamily arXiv:1303.4435 [hep-th]}}.

\bibitem{Kamada:2014qta}
K.~Kamada, Y.~Miyamoto, D.~Yamauchi, and J.~Yokoyama, ``{Effects of cosmic
  strings with delayed scaling on CMB anisotropy},''
  \href{http://dx.doi.org/10.1103/PhysRevD.90.083502}{{\em Phys. Rev. D}
  {\bfseries 90} no.~8, (2014) 083502},
  \href{http://arxiv.org/abs/1407.2951}{{\ttfamily arXiv:1407.2951
  [astro-ph.CO]}}.

\bibitem{Zhang:2015bga}
J.~Zhang, J.~J. Blanco-Pillado, J.~Garriga, and A.~Vilenkin, ``{Topological
  Defects from the Multiverse},''
  \href{http://dx.doi.org/10.1088/1475-7516/2015/05/059}{{\em JCAP} {\bfseries
  05} (2015) 059}, \href{http://arxiv.org/abs/1501.05397}{{\ttfamily
  arXiv:1501.05397 [hep-th]}}.

\bibitem{Ringeval:2015ywa}
C.~Ringeval, D.~Yamauchi, J.~Yokoyama, and F.~R. Bouchet, ``{Large scale CMB
  anomalies from thawing cosmic strings},''
  \href{http://dx.doi.org/10.1088/1475-7516/2016/02/033}{{\em JCAP} {\bfseries
  02} (2016) 033}, \href{http://arxiv.org/abs/1510.01916}{{\ttfamily
  arXiv:1510.01916 [astro-ph.CO]}}.

\bibitem{Guedes:2018afo}
G.~S.~F. Guedes, P.~P. Avelino, and L.~Sousa, ``{Signature of inflation in the
  stochastic gravitational wave background generated by cosmic string
  networks},'' \href{http://dx.doi.org/10.1103/PhysRevD.98.123505}{{\em Phys.
  Rev. D} {\bfseries 98} no.~12, (2018) 123505},
  \href{http://arxiv.org/abs/1809.10802}{{\ttfamily arXiv:1809.10802
  [astro-ph.CO]}}.

\bibitem{Cui:2019kkd}
Y.~Cui, M.~Lewicki, and D.~E. Morrissey, ``{Gravitational Wave Bursts as
  Harbingers of Cosmic Strings Diluted by Inflation},''
  \href{http://dx.doi.org/10.1103/PhysRevLett.125.211302}{{\em Phys. Rev.
  Lett.} {\bfseries 125} no.~21, (2020) 211302},
  \href{http://arxiv.org/abs/1912.08832}{{\ttfamily arXiv:1912.08832
  [hep-ph]}}.

\bibitem{Caldwell:1991jj}
R.~R. Caldwell and B.~Allen, ``{Cosmological constraints on cosmic string
  gravitational radiation},''
  \href{http://dx.doi.org/10.1103/PhysRevD.45.3447}{{\em Phys. Rev. D}
  {\bfseries 45} (1992) 3447--3468}.

\bibitem{Martins:1996jp}
C.~J. A.~P. Martins and E.~P.~S. Shellard, ``{Quantitative string evolution},''
  \href{http://dx.doi.org/10.1103/PhysRevD.54.2535}{{\em Phys. Rev. D}
  {\bfseries 54} (1996) 2535--2556},
  \href{http://arxiv.org/abs/hep-ph/9602271}{{\ttfamily arXiv:hep-ph/9602271}}.

\bibitem{Martins:2000cs}
C.~J. A.~P. Martins and E.~P.~S. Shellard, ``{Extending the velocity dependent
  one scale string evolution model},''
  \href{http://dx.doi.org/10.1103/PhysRevD.65.043514}{{\em Phys. Rev. D}
  {\bfseries 65} (2002) 043514},
  \href{http://arxiv.org/abs/hep-ph/0003298}{{\ttfamily arXiv:hep-ph/0003298}}.

\bibitem{Sousa:2013aaa}
L.~Sousa and P.~P. Avelino, ``{Stochastic Gravitational Wave Background
  generated by Cosmic String Networks: Velocity-Dependent One-Scale model
  versus Scale-Invariant Evolution},''
  \href{http://dx.doi.org/10.1103/PhysRevD.88.023516}{{\em Phys. Rev. D}
  {\bfseries 88} no.~2, (2013) 023516},
  \href{http://arxiv.org/abs/1304.2445}{{\ttfamily arXiv:1304.2445
  [astro-ph.CO]}}.

\bibitem{Sousa:2014gka}
L.~Sousa and P.~P. Avelino, ``{Stochastic gravitational wave background
  generated by cosmic string networks: The small-loop regime},''
  \href{http://dx.doi.org/10.1103/PhysRevD.89.083503}{{\em Phys. Rev. D}
  {\bfseries 89} no.~8, (2014) 083503},
  \href{http://arxiv.org/abs/1403.2621}{{\ttfamily arXiv:1403.2621
  [astro-ph.CO]}}.

\bibitem{Martins:2005es}
C.~J. A.~P. Martins and E.~P.~S. Shellard, ``{Fractal properties and
  small-scale structure of cosmic string networks},''
  \href{http://dx.doi.org/10.1103/PhysRevD.73.043515}{{\em Phys. Rev. D}
  {\bfseries 73} (2006) 043515},
  \href{http://arxiv.org/abs/astro-ph/0511792}{{\ttfamily
  arXiv:astro-ph/0511792}}.

\bibitem{Ringeval:2005kr}
C.~Ringeval, M.~Sakellariadou, and F.~Bouchet, ``{Cosmological evolution of
  cosmic string loops},''
  \href{http://dx.doi.org/10.1088/1475-7516/2007/02/023}{{\em JCAP} {\bfseries
  02} (2007) 023}, \href{http://arxiv.org/abs/astro-ph/0511646}{{\ttfamily
  arXiv:astro-ph/0511646}}.

\bibitem{Lorenz:2010sm}
L.~Lorenz, C.~Ringeval, and M.~Sakellariadou, ``{Cosmic string loop
  distribution on all length scales and at any redshift},''
  \href{http://dx.doi.org/10.1088/1475-7516/2010/10/003}{{\em JCAP} {\bfseries
  10} (2010) 003}, \href{http://arxiv.org/abs/1006.0931}{{\ttfamily
  arXiv:1006.0931 [astro-ph.CO]}}.

\bibitem{Blanco-Pillado:2011egf}
J.~J. Blanco-Pillado, K.~D. Olum, and B.~Shlaer, ``{Large parallel cosmic
  string simulations: New results on loop production},''
  \href{http://dx.doi.org/10.1103/PhysRevD.83.083514}{{\em Phys. Rev. D}
  {\bfseries 83} (2011) 083514},
  \href{http://arxiv.org/abs/1101.5173}{{\ttfamily arXiv:1101.5173
  [astro-ph.CO]}}.

\bibitem{Sanidas:2012ee}
S.~A. Sanidas, R.~A. Battye, and B.~W. Stappers, ``{Constraints on cosmic
  string tension imposed by the limit on the stochastic gravitational wave
  background from the European Pulsar Timing Array},''
  \href{http://dx.doi.org/10.1103/PhysRevD.85.122003}{{\em Phys. Rev. D}
  {\bfseries 85} (2012) 122003},
  \href{http://arxiv.org/abs/1201.2419}{{\ttfamily arXiv:1201.2419
  [astro-ph.CO]}}.

\bibitem{Blanco-Pillado:2013qja}
J.~J. Blanco-Pillado, K.~D. Olum, and B.~Shlaer, ``{The number of cosmic string
  loops},'' \href{http://dx.doi.org/10.1103/PhysRevD.89.023512}{{\em Phys. Rev.
  D} {\bfseries 89} no.~2, (2014) 023512},
  \href{http://arxiv.org/abs/1309.6637}{{\ttfamily arXiv:1309.6637
  [astro-ph.CO]}}.

\bibitem{Blanco-Pillado:2017oxo}
J.~J. Blanco-Pillado and K.~D. Olum, ``{Stochastic gravitational wave
  background from smoothed cosmic string loops},''
  \href{http://dx.doi.org/10.1103/PhysRevD.96.104046}{{\em Phys. Rev. D}
  {\bfseries 96} no.~10, (2017) 104046},
  \href{http://arxiv.org/abs/1709.02693}{{\ttfamily arXiv:1709.02693
  [astro-ph.CO]}}.

\bibitem{Blanco-Pillado:2017rnf}
J.~J. Blanco-Pillado, K.~D. Olum, and X.~Siemens, ``{New limits on cosmic
  strings from gravitational wave observation},''
  \href{http://dx.doi.org/10.1016/j.physletb.2018.01.050}{{\em Phys. Lett. B}
  {\bfseries 778} (2018) 392--396},
  \href{http://arxiv.org/abs/1709.02434}{{\ttfamily arXiv:1709.02434
  [astro-ph.CO]}}.

\bibitem{King:2021gmj}
S.~F. King, S.~Pascoli, J.~Turner, and Y.-L. Zhou, ``{Confronting SO(10) GUTs
  with proton decay and gravitational waves},''
  \href{http://dx.doi.org/10.1007/JHEP10(2021)225}{{\em JHEP} {\bfseries 10}
  (2021) 225}, \href{http://arxiv.org/abs/2106.15634}{{\ttfamily
  arXiv:2106.15634 [hep-ph]}}.

\bibitem{King:2023wkm}
S.~F. King, G.~K. Leontaris, and Y.-L. Zhou, ``{Flipped SU(5): unification,
  proton decay, fermion masses and gravitational waves},''
  \href{http://dx.doi.org/10.1007/JHEP03(2024)006}{{\em JHEP} {\bfseries 03}
  (2024) 006}, \href{http://arxiv.org/abs/2311.11857}{{\ttfamily
  arXiv:2311.11857 [hep-ph]}}.

\bibitem{Vilenkin:1991zk}
A.~Vilenkin, ``{Cosmic string dynamics with friction},''
  \href{http://dx.doi.org/10.1103/PhysRevD.43.1060}{{\em Phys. Rev. D}
  {\bfseries 43} (1991) 1060--1062}.

\bibitem{Kamada:2011bt}
K.~Kamada, K.~Nakayama, and J.~Yokoyama, ``{Phase transition and monopole
  production in supergravity inflation},''
  \href{http://dx.doi.org/10.1103/PhysRevD.85.043503}{{\em Phys. Rev. D}
  {\bfseries 85} (2012) 043503},
  \href{http://arxiv.org/abs/1110.3904}{{\ttfamily arXiv:1110.3904 [hep-ph]}}.

\bibitem{Damour:2001bk}
T.~Damour and A.~Vilenkin, ``{Gravitational wave bursts from cusps and kinks on
  cosmic strings},'' \href{http://dx.doi.org/10.1103/PhysRevD.64.064008}{{\em
  Phys. Rev. D} {\bfseries 64} (2001) 064008},
  \href{http://arxiv.org/abs/gr-qc/0104026}{{\ttfamily arXiv:gr-qc/0104026}}.

\bibitem{Binetruy:2009vt}
P.~Binetruy, A.~Bohe, T.~Hertog, and D.~A. Steer, ``{Gravitational Wave Bursts
  from Cosmic Superstrings with Y-junctions},''
  \href{http://dx.doi.org/10.1103/PhysRevD.80.123510}{{\em Phys. Rev. D}
  {\bfseries 80} (2009) 123510},
  \href{http://arxiv.org/abs/0907.4522}{{\ttfamily arXiv:0907.4522 [hep-th]}}.

\bibitem{2009ApJ...707..916F}
D.~J. {Fixsen}, ``{The Temperature of the Cosmic Microwave Background},''
  \href{http://dx.doi.org/10.1088/0004-637X/707/2/916}{{\em Astrophysical
  Journal} {\bfseries 707} no.~2, (Dec., 2009) 916--920},
  \href{http://arxiv.org/abs/0911.1955}{{\ttfamily arXiv:0911.1955
  [astro-ph.CO]}}.

\bibitem{Planck:2018vyg}
{\bfseries Planck} Collaboration, N.~Aghanim {\em et~al.}, ``{Planck 2018
  results. VI. Cosmological parameters},''
  \href{http://dx.doi.org/10.1051/0004-6361/201833910}{{\em Astron. Astrophys.}
  {\bfseries 641} (2020) A6}, \href{http://arxiv.org/abs/1807.06209}{{\ttfamily
  arXiv:1807.06209 [astro-ph.CO]}}. [Erratum: Astron.Astrophys. 652, C4
  (2021)].

\bibitem{Schmitz:2024hxw}
K.~Schmitz and T.~Schr\"oder, ``{Gravitational waves from low-scale cosmic
  strings},'' \href{http://dx.doi.org/10.1103/PhysRevD.110.063549}{{\em Phys.
  Rev. D} {\bfseries 110} no.~6, (2024) 063549},
  \href{http://arxiv.org/abs/2405.10937}{{\ttfamily arXiv:2405.10937
  [astro-ph.CO]}}.

\bibitem{Janssen:2014dka}
G.~Janssen {\em et~al.}, ``{Gravitational wave astronomy with the SKA},''
  \href{http://dx.doi.org/10.22323/1.215.0037}{{\em PoS} {\bfseries AASKA14}
  (2015) 037}, \href{http://arxiv.org/abs/1501.00127}{{\ttfamily
  arXiv:1501.00127 [astro-ph.IM]}}.

\bibitem{Kawamura:2020pcg}
S.~Kawamura {\em et~al.}, ``{Current status of space gravitational wave antenna
  DECIGO and B-DECIGO},'' \href{http://dx.doi.org/10.1093/ptep/ptab019}{{\em
  PTEP} {\bfseries 2021} no.~5, (2021) 05A105},
  \href{http://arxiv.org/abs/2006.13545}{{\ttfamily arXiv:2006.13545 [gr-qc]}}.

\bibitem{Harry:2006fi}
G.~M. Harry, P.~Fritschel, D.~A. Shaddock, W.~Folkner, and E.~S. Phinney,
  ``{Laser interferometry for the big bang observer},''
  \href{http://dx.doi.org/10.1088/0264-9381/23/15/008}{{\em Class. Quant.
  Grav.} {\bfseries 23} (2006) 4887--4894}. [Erratum: Class.Quant.Grav. 23,
  7361 (2006)].

\bibitem{LIGOScientific:2016wof}
{\bfseries LIGO Scientific} Collaboration, B.~P. Abbott {\em et~al.},
  ``{Exploring the Sensitivity of Next Generation Gravitational Wave
  Detectors},'' \href{http://dx.doi.org/10.1088/1361-6382/aa51f4}{{\em Class.
  Quant. Grav.} {\bfseries 34} no.~4, (2017) 044001},
  \href{http://arxiv.org/abs/1607.08697}{{\ttfamily arXiv:1607.08697
  [astro-ph.IM]}}.

\bibitem{Hild:2008ng}
S.~Hild, S.~Chelkowski, and A.~Freise, ``{Pushing towards the ET sensitivity
  using 'conventional' technology},''
  \href{http://arxiv.org/abs/0810.0604}{{\ttfamily arXiv:0810.0604 [gr-qc]}}.

\bibitem{Fu:2024rsm}
B.~Fu, A.~Ghoshal, S.~F. King, and M.~H. Rahat, ``{Type-I two-Higgs-doublet
  model and gravitational waves from domain walls bounded by strings},''
  \href{http://dx.doi.org/10.1007/JHEP08(2024)237}{{\em JHEP} {\bfseries 08}
  (2024) 237}, \href{http://arxiv.org/abs/2404.16931}{{\ttfamily
  arXiv:2404.16931 [hep-ph]}}.
  \url{https://github.com/moinulrahat/CosmicStringGW/}.

\bibitem{Dine:1995uk}
M.~Dine, L.~Randall, and S.~D. Thomas, ``{Supersymmetry breaking in the early
  universe},'' \href{http://dx.doi.org/10.1103/PhysRevLett.75.398}{{\em Phys.
  Rev. Lett.} {\bfseries 75} (1995) 398--401},
  \href{http://arxiv.org/abs/hep-ph/9503303}{{\ttfamily arXiv:hep-ph/9503303}}.

\bibitem{Dine:1995kz}
M.~Dine, L.~Randall, and S.~D. Thomas, ``{Baryogenesis from flat directions of
  the supersymmetric standard model},''
  \href{http://dx.doi.org/10.1016/0550-3213(95)00538-2}{{\em Nucl. Phys. B}
  {\bfseries 458} (1996) 291--326},
  \href{http://arxiv.org/abs/hep-ph/9507453}{{\ttfamily arXiv:hep-ph/9507453}}.

\bibitem{Allen:1983dg}
B.~Allen, ``{Phase Transitions in de Sitter Space},''
  \href{http://dx.doi.org/10.1016/0550-3213(83)90470-4}{{\em Nucl. Phys. B}
  {\bfseries 226} (1983) 228--252}.

\bibitem{Ishikawa:1983kz}
K.~Ishikawa, ``{GRAVITATIONAL EFFECT ON EFFECTIVE POTENTIAL},''
  \href{http://dx.doi.org/10.1103/PhysRevD.28.2445}{{\em Phys. Rev. D}
  {\bfseries 28} (1983) 2445}.

\bibitem{Starobinsky:1980te}
A.~A. Starobinsky, ``{A New Type of Isotropic Cosmological Models Without
  Singularity},'' \href{http://dx.doi.org/10.1016/0370-2693(80)90670-X}{{\em
  Phys. Lett. B} {\bfseries 91} (1980) 99--102}.

\bibitem{Barrow:1988xh}
J.~D. Barrow and S.~Cotsakis, ``{Inflation and the Conformal Structure of
  Higher Order Gravity Theories},''
  \href{http://dx.doi.org/10.1016/0370-2693(88)90110-4}{{\em Phys. Lett. B}
  {\bfseries 214} (1988) 515--518}.

\end{thebibliography}\endgroup

\end{document}